\documentclass[aps,pra,twocolumn,showpacs,amsmath,amssymb,floatfix,superscriptaddress,notitlepage,nofootinbib]{revtex4-1}
\usepackage{amsmath, mathrsfs}
\usepackage{amsthm}
\usepackage{amssymb}
\usepackage{amsfonts}
\usepackage{bbm}
\usepackage{epsfig}
\usepackage{times}
\usepackage{xcolor}
\usepackage{hyperref}
\usepackage{framed}

\hypersetup{
	colorlinks=true,
	linkcolor=blue,
	citecolor=blue,
	urlcolor=blue
}

\usepackage[justification=justified]{caption}
\usepackage{subcaption}

\newcommand{\tr}{\text{Tr}}
\newcommand{\new}[1]{ #1}

\newcommand{\supp}[1]{\text{supp}(#1)}
\newcommand{\norm}[1]{\vert \vert #1 \vert \vert}

\begin{document}
	
	\title{Quantum many-body systems in thermal equilibrium}
	
	\author{\'Alvaro M. Alhambra }
		\email{alvaro.alhambra@csic.es}
	\affiliation{ Max-Planck-Institut f\" ur Quantenoptik, Hans-Kopfermann-Strasse 1, D-85748 Garching, Germany}
	\affiliation{
Instituto de F\'isica Te\'orica UAM/CSIC, C/ Nicol\'as Cabrera 13-15, Cantoblanco, 28049 Madrid, Spain}

	\begin{abstract}
	The thermal or equilibrium ensemble is one of the most ubiquitous states of matter. For models comprised of many locally interacting quantum particles, it describes a wide range of physical situations, relevant to condensed matter physics, high energy physics, quantum chemistry and quantum computing, among others. We give a pedagogical overview of some of the most important universal features about the physics and complexity of these states, which have the locality of the Hamiltonian at its core. We focus on mathematically rigorous statements, many of them inspired by ideas and tools from quantum information theory. These include bounds on their correlations, the form of the subsystems, various statistical properties, and the performance of classical and quantum algorithms. We also include a summary of a few of the most important technical tools, as well as some self-contained proofs. 
	\end{abstract}
	
		\maketitle
	
		\tableofcontents
	
%\vspace{10pt}
%\noindent\rule{\textwidth}{1pt}
%\tableofcontents%\thispagestyle{fancy}
%\noindent\rule{\textwidth}{1pt}
%\vspace{10pt}

\section{Introduction}

We are currently at the dawn of the age of synthetic quantum matter. Increasingly better experiments on a variety of quantum platforms are improving in size and controllability at unprecedented rates, aided by the current impulse of quantum information science and technology. This gives very good prospects to the exploration of the physics of complex quantum many body systems. Our aspiration to better understand these systems is very well motivated from a scientific perspective, but also potentially from the industrial one: unlocking the potential of complex quantum systems may bring surprising advances to the engineering of new materials or chemical compounds in the future. It may also yield computational tools with unprecedented capabilities for a still unknown range of applications.

Many of the most commonly studied materials and current experimental platforms are described by an arrangement of quantum particles in some sort of lattice configuration. Due to the spatial decay of electromagnetic forces, each of these particles only interacts appreciably with their immediate vicinity, which causes the interactions between them to be local. 

In this tutorial, we focus on the properties of these important systems when at thermal equilibrium, so that they are accurately described by the so-called thermal or Gibbs state. We review and explain some of their most important universal properties, covered from a mathematical perspective. That is, we focus on statements that can be proven about states of the Gibbs form
\begin{equation}
    \rho_\beta = \frac{e^{-\beta H}}{ Z}= \frac{1}{Z}\sum_l e^{-\beta E_l} \vert E_l \rangle \langle E_l \vert,
\end{equation}
where $H=\sum_l  E_l \vert E_l \rangle \langle E_l \vert$ is the Hamiltonian, $\beta$ is the inverse temperature and $Z \equiv \text{Tr} [e^{-\beta H}]$ is the partition function. The Hamiltonian describes the interactions between the $N$ particles, which are restricted to short-ranged or local. A ``local Hamiltonian" is a Hermitian operator $H$ in the finite-dimensional Hilbert space of $N$ $d$-dimensional particles $ (\mathbb{C}^{d})^{\otimes N }$. It is defined as a sum of terms 
\begin{equation}\label{eq:hamiltonian}
    H=\sum_i h_i \otimes \mathbb{I},
\end{equation}
each of which has support (i.e. acts non-trivially) on at most $k$ particles, and bounded strength, such that 
\begin{equation}\label{eq:h}
\max_i \vert \vert h_i \vert \vert =h
\end{equation}
For a definition of the operator norm $\vert \vert \cdot \vert \vert$ see Sec. \ref{sec:norms} below. \new{Typically, the Hamiltonians are scaled so that $h=\mathcal{O}(1)$.}

In what follows we just write the terms as $h_i$ for simplicity. These constitute the individual interactions, which are typically arranged in a lattice of a small dimension. A simple example is e.g. the transverse-field Ising model in one dimension  with open boundary conditions
\begin{equation}
    H_{\text{Ising}} =   \sum_{j=1}^{N-1} \left(J\sigma^{X}_j \sigma^{X}_{j+1} + \Delta \sigma^{Z}_j \right)+ \Delta \sigma^{Z}_N .
\end{equation}
Here, $k=2$ and the interactions are arranged on a 1D chain.

The idea of a local Hamiltonian is very general, and involves many different models describing a wide range of situations, of interest for many fields of physics, chemistry and computer science. The only thing they have in common is the \emph{locality} of the interactions. We aim to understand mathematically how this fact alone constrains both the physics and the computational complexity when combined with thermal fluctuations.

The thermal states of these general local Hamiltonians \new{appear in many different contexts, and} are interesting for a wide variety of reasons. Some of the main ones are:

\begin{itemize}
    \item It is one of the most \textbf{ubiquitous} states of quantum matter: typical experiments happen at finite temperature, where the quantum system at hand is weakly coupled to some external radiation field \new{or phonon bath}, that drive it to the thermal state. \new{For completeness, we sketch the standard argument of how the weak coupling assumption leads to states of the Gibbs form in App. \ref{app:weak}.}
    
    \item The thermal state is also important when studying not just systems with an external bath, but also in the evolution of \textbf{isolated} quantum systems, even when their global state is pure: in very generic cases, these end up being ``their own bath", and the individual subsystems thermalize to the Gibbs ensemble \cite{gogolin2016,dalessio2016}.
    
    \item From a general condensed matter/material science standpoint, we are very interested in numerous questions about the physics at \textbf{finite temperature}: How are conserved quantities (e.g. charge, energy) propagated in a state close to equilibrium? How does the system respond to small or large perturbations away from equilibrium? 
    
    \item Systems at thermal equilibrium (both quantum and classical) display interesting \textbf{phase transitions} in certain (low) temperature regimes (e.g. classical Ising model in 2D). It is thus relevant to study what are their universal properties both in and away from the critical points.
    
    \item They are also important from the point of view of \textbf{quantum phases of matter} and topological order \new{at zero temperature}. It has been widely established that thermal states of local models in dimension $D-1$ appear in the entanglement spectrum of $D$- dimensional ground states \cite{li-haldane2008}. As such, understanding their structure should also help us in elucidating the low energy behaviour of many interesting systems.
    
    \new{\item They very naturally appear in information theory and inference as the distributions that best reproduce partial current knowledge of a system. This is justified by Jaynes' principle,  which we explain in App. \ref{app:jaynes}.}
    
    \item These states are also important for \textbf{computation}. For instance, being able to sample from the thermal distribution of local models is a typical subroutine for certain classical and quantum algorithms \cite{BrandaoSvore2017,brando_et_al:LIPIcs:2019:10603,vanApeldoorn2020quantumsdpsolvers,GSLBrandao2022fasterquantum}. They are also a very naturally occurring data structure in both classical and quantum machine learning \cite{PhysRevX.8.021050,PhysRevA.96.062327,Biamonte2017,BaireySampling2019,Anshu_2021} (often under the name of \emph{Boltzmann machines}).
    
    \item It is known from \textbf{quantum computational complexity} that the low energy subspace of local Hamiltonians is able to encode the solution to very hard computational problems: finding the ground state energy is \textbf{QMA complete} \cite{QMA2005}. Thus, it is widely believed that even a quantum computer should not be able to do it in polynomial time. This then at least also applies to the thermal state at very low temperature, and motivates the study of how the complexity changes as the temperature rises \cite{AharonovPCP2013,HastingsPCP2013}.

\end{itemize}

There are many different specific aspects that one could explore, but here we focus on the following, which we believe to be of particular importance:
\begin{itemize}
    \item The correlations between the particles at different parts of the lattice. \new{In particular, how does the structure of those correlations are structured in relation to the geometry of the lattice.}
    \item The \new{states} of the subsystems that a thermal state can take, and how are they related to few-particle Gibbs states.
     \item The statistical physics properties of these systems at equilibrium, including Jaynes' principle, concentration bounds and equivalence of ensembles.
    \item The efficiency of classical and quantum algorithms for the generation and manipulation of thermal states, and the computation of expectation values and partition functions.
\end{itemize}
\new{More specifically, we focus on these topics for a broad family of Gibbs states that can be understood as being away from phase transitions within the phase space. It is for this region of the parameter space of Hamiltonians that the mathematical results described here are typically more tractable and give insightful results. We describe this further in Sec. \ref{sec:completely}.}

The general topic of this tutorial, and the particular results explained here, are a small part of the exciting past, present and future efforts to understand the physics and complexity of quantum many-body systems. We hope to contribute to the understanding and cross-fertilization of the many different angles that the quantum many-body problem can take. See also e.g. \cite{Kliesch2018,gogolin2016} for previous references with partially overlapping content.

\subsection{Scope and content}

Throughout this tutorial, we cover statements that have a precise mathematical formulation, many of them motivated by a quantum information theoretic perspective. This notably includes a short exposition of a few key technical tools in Sec. \ref{sec:tools}. These have not previously appeared together, but rather separately explained in the literature with various levels of detail, depending on the context and usage. We hope that this encourages new, potentially unexpected, applications thereof.

Along the sections with the actual physical and computational results, we write the proofs of some of the simpler or more important ones explained throughout. This includes at least one main result per section, which should serve as a pedagogical example. For the rest, some of which have more detailed or involved derivations, we refer the reader to the original works cited along the text. One of our main hopes is that after reading this tutorial even the more technical works will be more easily accessible to a wider range of researchers. Because of this, rather than the traditional Theorem-Proof structure of most mathematical physics writing, we have chosen a more streamlined style for the presentation which allows for more physical explanations and intuitions of the steps. This will hopefully contribute to a wider readability.

\new{Most of what we describe are known results, with at most some small improvements}, with the proofs either being the same or slightly simplified versions of previous ones. The relevant references are included, but this does not mean that all of the previous relevant ones are listed here: we are certainly missing to mention a very large body of work. This includes many relevant papers on mathematical physics, but also a lot of important physics literature that covers these topics from perspectives that are beyond our scope: based on numerical methods, theory work on experimental implementations, as well as all experimental results. 

%%%%%%%
%%%%%%%

\subsection{Completely analytical interactions}\label{sec:completely}

\new{
Before we proceed, we should put the content of this tutorial into context more precisely. In the mathematical study of many body systems, one of the biggest points of interest are phase transitions, including the study of order parameters, symmetry breaking, and other very well established ideas which aim at classifying the possible kinds of phase transitions. For instance, there are important models, such as the paradigmatic Ising in two and three dimensions, that have a very well understood phase transition at a given temperature.}

\new{However, for  many classes of models and most regions of their parameter space, local Hamiltonians do not have a thermal phase transition. This ``one phase region" (as it is sometimes referred to \cite{Lenci2005}) likely contains the ``simplest" cases of thermal equilibrium, including those of non-interacting gases. These situations are typically characterized by the analiticity of the partition function and other closely related facts, such as:
\begin{itemize}
 \item The convergence of the cluster expansion.
    \item The localization of correlations in the lattice, and absence of long range order.
     \item The approximation of marginals with local Gibbs states, and the idea of locality of temperature.
    \item Concentration properties of local observables.
    \item Efficiency of approximation, either with quantum or classical algorithms.
    \item The existence and boundedness of log-Sobolev constants.
\end{itemize}}

\new{It is expected that many or all of these simplifying facts are equivalent, in that a model that obeys one (such as the analiticity of the partition function) will also display the other features.
In the classical case, a large number of conditions are known to be equivalent to the analiticity of the partition function. The study of this problem was initiated by the seminal work of Dobrushing and Shlosman \cite{Dobrushin1987}, aiming at characterizing these ``completely analytical interactions" in terms of many diferent equivalent conditions (12 in the original article). In the quantum case, much less is known about the equivalence of the analiticity of the partition function with other physical facts, although some important steps have been taken (see e.g. \cite{Harrow_2020,capel2020modified}).}

\new{The main aim of this tutorial is to cover results that show that Gibbs states have simplifying features with respect to generic quantum states. Following that spirit, most (although not all) of the results explained here apply to this ``phase" or universality class of Gibbs states, in which those simplifying facts are expected to hold. In fact, every element of the list above is individually considered in each of the sections below. Because of that, we do not cover an important part of the literature where analytical results are typically much harder to obtain. For instance, those studying the effect of phase transitions in e.g. the simulability of Gibbs states, the types of correlations that can arise, and others. }

%%%%%%%%
%%%%%%%%

\section{Mathematical preliminaries and notation}

\subsection{Operator norms}\label{sec:norms}

A basic but very important mathematical tool in this context are the Schatten $p$-norms for operators, as well as the different inequalities between them. These norms are maps from the space of operators to $\mathbb{R}$, as $M \rightarrow \vert \vert M \vert \vert_p$, that obey the following properties:
\begin{itemize}
    \item Homogeneous: If $\alpha$ is a scalar, $\vert \vert  \alpha M \vert \vert_p = \vert \alpha \vert \vert \vert M \vert \vert_p$.
    \item Positive: $\vert \vert M \vert \vert_p \ge 0$.
    \item Definite: $\vert \vert M \vert \vert_p =0 \iff M=0$.
    \item Triangle inequality: $\vert \vert M_1+M_2 \vert \vert_p \le  \vert \vert M_1 \vert \vert_p + \vert \vert M_2 \vert \vert_p $.
\end{itemize}
For a given operator $M$ with singular values $\{\alpha_l^M\}$ and $p\in [1,\infty)$, they are defined as
\begin{equation} 
    \vert \vert M \vert \vert_p \equiv \text{Tr}[ \vert M \vert^p]^{\frac{1}{p}} = \left( \sum_l (\alpha_l^M)^p \right)^{\frac{1}{p}}.
\end{equation}
The more important ones are the operator norm $\vert \vert M \vert \vert \equiv \vert \vert M \vert \vert_{\infty} = \max_l \vert \alpha_l^M \vert$, the Hilbert-Schmidt $2$-norm $\vert \vert M \vert \vert_2=\text{Tr}[M M^\dagger]^{1/2}$ and the 1-norm or trace norm
\begin{equation} \label{eq:1norm}
    \vert \vert M \vert \vert_1 = \max_{\vert \vert P \vert \vert \le 1}\text{Tr}[M P].
\end{equation}
Thus $\vert \tr[M]\vert \le \vert \vert M \vert \vert_1$, with equality for positive operators. For quantum states, $\tr[\rho]=\vert \vert \rho \vert \vert_1=1$.

Typically we measure the ``strength" of an observable with the operator norm, and the closeness of two quantum states with the trace norm $\vert \vert \rho - \sigma \vert \vert_1$, since it is related to the probability of distinguishing them under measurements. The $2$-norm, on the other hand, is often the easiest one to compute in practice. Also note the very important H\"older's inequality
\begin{equation}\label{eq:Holder}
    \vert \vert M_1 M_2 \vert \vert_p \le \vert \vert M_1 \vert \vert_{q_1}  \vert \vert M_2 \vert \vert_{q_2}, 
\end{equation}
which holds for $\frac{1}{p}=\frac{1}{q_1}+\frac{1}{q_2}$ (e.g. $p=q_1=1,q_2=\infty$). A particularly useful corollary is the Cauchy-Schwarz inequality, when $q_1=q_2=2$ and $p=1$,
\begin{equation}
    \vert \tr[M_1^\dagger M_2 ]\vert^2 \le \tr[M_1 M_1^\dagger] \tr[M_2 M_2^\dagger].
\end{equation}

\subsection{Information-theoretic quantities}
Let us define the von Neumann entropy of a quantum state $\rho$ \footnote{The $\log$s here are defined with base $e$.}
\begin{equation}
S(\rho)= -\text{Tr}[\rho \log (\rho)],
\end{equation}
which, roughly speaking, quantifies the uncertainty we have about the particular state. It is bounded by $0 \le S(\rho) \le \log d$. The lower bound is obtained by choosing $\rho$ pure, and the upper bound by the identity $\rho = \mathbb{I}/d$.  Another important quantity is the Umegaki relative entropy
\begin{equation}
    D(\rho \vert \sigma ) = \tr[\rho (\log \rho - \log \sigma)],
\end{equation}
which is a measure of distinguishability of quantum states.  It obeys Pinsker's inequality \begin{equation}
D(\rho \vert \sigma ) \ge \frac{1}{2}\vert \vert \rho-\sigma \vert \vert_1^2,
\end{equation} which relates the relative entropy with the 1-norm. It is strictly positive for $\rho \neq \sigma$, and vanishes otherwise. It is also closely related to the non-equilibrium free energy
\begin{equation}\label{eq:relFree}
    D(\rho \vert \rho_\beta ) = \beta \tr[\rho H] - S(\rho) + \log Z \equiv \beta F_\beta(\rho) + \log Z,
\end{equation}
which also shows that the equilibrium free energy is $F_\beta (\rho_\beta)=-\beta ^{-1}\log Z$. \new{This distance measure also naturally appears in the derivation of Jayne's principle, as shown in App. \ref{app:jaynes}.}

From these quantities we can also define the quantum mutual information, which, given a bipartite state $\rho^{AB}$ on subsystems $A$ and $B$, with $\tr_{B}[\rho^{AB}]=\rho_A, \tr_{A}[\rho^{AB}]=\rho^B$, quantifies the correlations between $A$ and $B$ as
\begin{align}\label{eq:MI}
    I(A:B)_{\rho}&= S(\rho^A)+S(\rho^B) - S(\rho^{AB} ) \\&= D (\rho^{AB}\vert \rho^A \otimes \rho^B).
\end{align}
\new{In particular,} it is zero if and only if $\rho^{AB} = \rho^A \otimes \rho^B$. For all these three functions we can also define their corresponding R\'enyi generalizations. See \cite{khatri2020,scalet2021computable} for details.

A further, perhaps more refined quantity is the \emph{conditional mutual information} (CMI), defined as
\begin{align}
I(A:C \vert B)_\rho= & S(\rho^{AB})+S(\rho^{BC})-S(\rho^{ABC})-S(\rho^{B})\nonumber
\\ = & I(A:BC)_{\rho}-I(A:B)_{\rho}.\label{eq:CMI}
\end{align}
This perhaps less known quantity is behind many non-trivial statements in quantum information theory (see Section 11.7 in \cite{WildeBook} for more details). In a nutshell, it measures how much $A$ and $C$ share correlations that are \emph{not} mediated by $B$. That is, if this quantity is small, most of the correlations between $A$ and $C$ (which may be weak) are in reality correlations between $A$ and $B$ and $B$ and $C$. 

\subsection{Lattice notation}\label{sec:latticenotation}

In what follows we need some technical definitions regarding the properties of the Hamiltonian and the lattice. The lattice is a hypergraph which we denote by $\Lambda=\{V,E\}$ with vertex set $V$ and hyperedges $E$. To each vertex we associate a Hilbert space of dimension $d$, $\mathbb{C}^d$. The number of particles is $N= \vert V \vert$, and the number of hyperedges is $\vert E \vert$. The locality of the Hamiltonian can be expressed by a parameter $\frak{d}$, defined as the as the largest number of hyperedges adjacent to any individual hyperedge.

We can separate the vertices into subregions, such as $V_A$, and we denote with $\partial_A \in V_A$ the sites at the boundary of that region (that is, with at least one hyperedge connecting to $\setminus \Lambda_A$), of which there are $\vert \partial_A \vert$. For simplicity, we often refer to regions as $A,B,..$ instead of $V_A, V_B,...$. We also need the notion of ``distance'' between two regions, $\text{dist}(A,B)$, defined as the smallest number of overlapping hyperedges that connect a vertex from $A$ with a vertex from $B$.

To define the Hamiltonian, we associate local interactions to hyperedges, such that $H=\sum_{i \in E} h_i$. For an operator $h_i$, the set of vertices on which it has non-trivial support is $\text{supp}(h_i)$. We have already specified that each $h_i$ is such that $\vert \supp{h_i} \vert \le k$ (that is, the hyperedges have size at most k), so that for constant $k$, $N \propto \vert E \vert$. We also note that that $\vert \vert h_i \vert \vert \le h$ and introduce the following quantity
\begin{equation}
    J = \max_{x \in V} \sum_{i: x \in \text{supp}(h_i)} \vert \vert h_i \vert \vert,
\end{equation}
that is, $J$ upper bounds the norm of the interactions that act on an individual vertex.  

\subsection{Asymptotic  notation}\label{sec:bigO}

The so-called asymptotic or Bachmann–Landau notation succinctly describes the asymptotic behaviour of a function when the argument grows large. It is typically used when in a particular expression there are constant factors that we are happy to omit, that are unnecessarily cumbersome, or when we only have partial knowledge of the \new{exact expression but know the asymptotic behaviour}. 
We say that , given functions $f(N),g(N) \ge 0$:
\begin{itemize}
    \item $f(N)= \mathcal{O}(g(N))$ if there are constants $M,N_0>0$ such that $\forall N > N_0$, $ f(N) \le M g(N)$.
     \item $f(N)= \tilde{\mathcal{O}}(g(N))$ is similar to $\mathcal{O}(g(N))$ but with possible additional poly-logarithmic factors, so that instead $\forall N > N_0$, $ f(N) \le M g(N) \text{polylog}(g(N))$.
    \item $f(N)=o(g(N)$ if for every $\varepsilon >0$ there exists a $N_0>0$ such that $\forall N > N_0$, $ f(N) \le  \epsilon g(N) $.
    \item $f(N)= \Omega(g(N))$ if there are constants $M,N_0>0$ such that $\forall N > N_0$, $ f(N) \ge  M g(N)$.
\end{itemize}
These are the most commonly used symbols of this notation, all of which appear below. 

\section{An overview of technical tools}\label{sec:tools}

When studying quantum thermal states from a mathematical point of view, what we often need is some way of simplifying the operator $e^{-\beta H}$, in a way that makes the particular problem at hand mathematically tractable. This is usually achieved by expressing the relevant function of $e^{-\beta H}$ in simpler terms. Potential issues that complicate this are:
\begin{enumerate}
    \item The exponential of a local operator is not a local operator, due to the high order terms in the expansion, and could in principle be arbitrarily complicated. 
    \item The individual elements in the Hamiltonian Eq. \eqref{eq:hamiltonian} do not commute with each other. Thus we cannot divide the exponential of the Hamiltonian into smaller pieces by iterating simple identities like $e^{-\beta(H_1+H_2)}\stackrel{?}{=} e^{-\beta H_1} e^{-\beta H_2}$.
\end{enumerate}

The locality of the Hamiltonian helps make these two problems often not as serious as they could be in general situations. There is a number of tools to deal with this, and we now describe some of the most relevant ones. Below, we explain how the cluster expansion in Sec. \ref{sec:connectedclusters} helps with issue 1, while there are at least two different techniques in Sec. \ref{sec:localityestimates} and \ref{sec:QBP} that help us with issue 2. 

%%%%%%%
%%%%%%%

\subsection{Connected cluster expansion}\label{sec:connectedclusters}

This is a powerful set of ideas whose origins can be traced back to a wide set of the classic (and classical) literature on mathematical physics and statistical mechanics (see e.g. \cite{ruelle1999statistical})\new{, initiated in \cite{Mayer1941}. It has traditionally been used to prove the analyticity of the partition function at high temperatures and other regimes, so it serves as an ideal tool to characterize the completely analytical interactions from Sec. \ref{sec:completely}. More recently, it has also been used to study the existence of computationally efficient approximation schemes to it (see e.g. \cite{Mann2021}). The technique is flexible and general enough that it can also cover objects beyond partition functions, such as characteristic functions and other related quantities. }

\new{For simplicity we here focus on the high temperature expansion \footnote{This type of expansion is not limited to high temperatures, it can be used to expand around any other parameter, such as a local magnetization. See e.g. Chapter 5 of  \cite{friedli2017}. }.} The starting point is the logarithm of the partition function $\log Z \equiv  \log \tr[e^{-\beta H}]$. Let us consider its Taylor expansion around $\beta=0$ 
\begin{equation} \label{eq:logZ}
\log Z=\sum_m \frac{\beta^m}{m!}K_m.
\end{equation}

One can then ask: what is the radius of convergence of this Taylor series? More precisely, we would like to know whether there is some $\beta^*$ \new{independent of the system size} such that for $0 \le \beta < \beta^*$ we have that:
\begin{itemize}
    \item The function $\log Z$ is analytic.
      \item The $m$-th derivative at $\beta=0$ is such that
      \begin{equation} \label{eq:Kmsmall}
          \left \vert \frac{\text{d}^m \log Z}{\text{d} \beta^m} \right \vert = \vert K_m \vert \le  C_1 N (\beta/\beta^*)^{m} m!,
      \end{equation}
      for some constant $C_1$.
    \item The truncated Taylor series gives a good approximation as
    \begin{equation}\label{eq:clusterconvergence}
        \left \vert \log Z - \sum_{m=0}^{M} \frac{\beta^m}{m!}K_m  \right \vert \le C_1 N \frac{(\beta/\beta^*)^{(M+1)}}{1-(\beta/\beta^*)}.
    \end{equation}
  
\end{itemize}

\new{There are various ways to narrow down the radius of convergence of this series, but they all revolve around the idea of writing $\log Z$ in terms of connected clusters.}

\begin{figure}%[ht]

        \includegraphics[trim=00 00 00 00,width=0.22\textwidth]{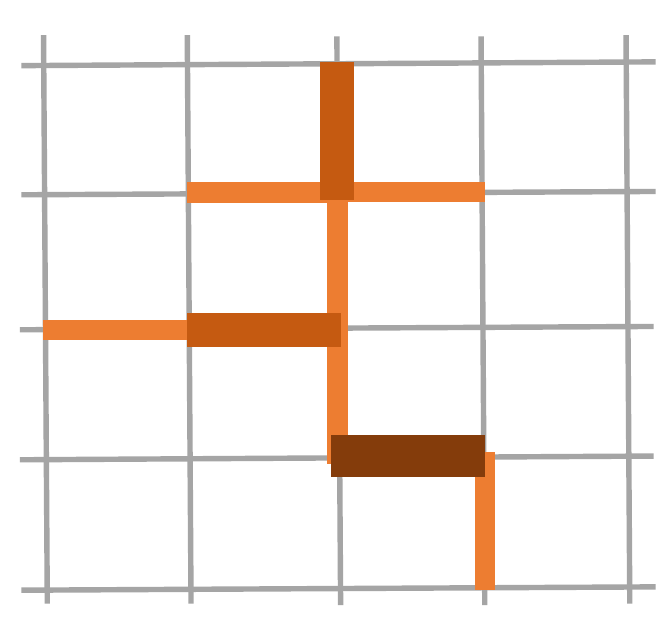} 
          \includegraphics[trim=00 00 00 00,width=0.22\textwidth]{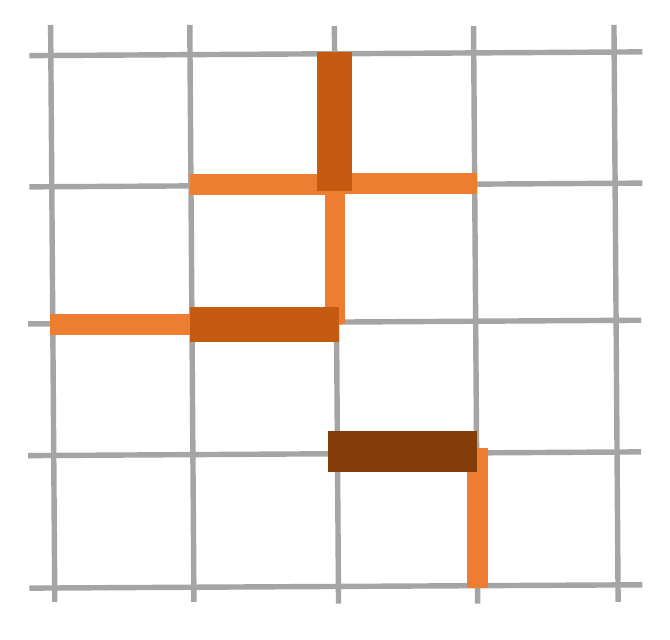}  
\caption{
\label{fig:clusters} 
\new{Illustration of the clusters, defined as a multiset of interaction terms or hyperedges. In this example, the interaction is a graph on a square lattice. The cluster on the left is connected $\textbf{W} \in \mathcal{G}_m$, while the one on the right is disconnected. The thickness of the lines represents the multiplicities $\mu_i^{\textbf{W}}$ of the edges, which may appear any number of times in a cluster as long as $\sum_i \mu_i^{\textbf{W}} = m$.}
}
\end{figure}

A cluster is a multiset (that is, a set counting multiplicities) of Hamiltonian terms $h_i$ (or alternatively, of hyperedges $\{i \in E\}$), which can appear more than once. A given cluster $\textbf{W}$ has size $\vert \textbf{W}\vert $ equal to the number of elements in the multiset (counting multiplicities $\mu^{\textbf{W}}_i$, so that $\vert \textbf{W}\vert=\sum_{\{i \in \textbf{W}\}}\mu^\textbf{W}_i$). Moreover, $\textbf{W}$ is connected if the hypergraph with hyperedges $i \in \textbf{W}$ is connected. Let us define the set of all clusters of size at most $\vert \textbf{W} \vert = m$ with $\mathcal{C}_m$, and the set of all connected clusters as $\mathcal{G}_m$. \new{For instance, $\mathcal{G}_1$ is the set of $\{h_i\}$, $\mathcal{G}_2$ are the pairs $\{ h_i , h_j \}$ provided that $i,j$ are adjacent or $i=j$ (in which case $\mu_i^{\textbf{W}}=2$). See Fig. \ref{fig:clusters} further illustrations of a connected and a disconnected cluster.}

Now, let us define the Hamiltonian with auxiliary variables $\{\lambda_i\}$ as $H(\lambda)=\sum_i \lambda_i h_i$. We use this to introduce the cluster derivative
\begin{equation}
    \mathcal{D}_{\textbf{W}} = \left. \prod_{i \in \textbf{W}} \left( \frac{\partial}{\partial \lambda_i} \right)^{\mu^{\textbf{W}}_i} \right|_{\lambda = 0}.
\end{equation}
Here, the subscript $\lambda = 0$ means to set $\lambda_i = 0$ for all $i$ after taking the derivatives. We thus write
\begin{equation} \label{eq:Kmoments}
\beta^m K_m = \sum_{\textbf{W} \in \mathcal{C}_m} \mathcal{D}_{\textbf{W}} \log \tr[e^{-\beta H(\lambda)}].
\end{equation}
 What we have done here is to simply write each moment of the Taylor series as a sum of the contributions of all clusters $\textbf{W}$, without further specifying what each contribution looks like.

We now prove the key simplification stemming from this expression. Let $\textbf{W} \notin \mathcal{G}_m$, so that we have $\textbf{W}= \textbf{W}_1 \cup \textbf{W}_2$ where $\textbf{W}_1, \textbf{W}_2$ are non-overlapping clusters. This allowsus to define  $h_{\textbf{W}_1},h_{\textbf{W}_2}$ as the Hamiltonian terms in those clusters, so that $\supp{h_{\textbf{W}_1}} \cap \supp h_{\textbf{W}_2}= \O$. We then have
\begin{align}
    \mathcal{D}_{\textbf{W}}& \log \tr[e^{-\beta H(\lambda)}]=\mathcal{D}_{\textbf{W}} \log \tr[e^{-\beta (h_{\textbf{W}_1}(\lambda)+h_{\textbf{W}_2}(\lambda))}] \nonumber \\ &= \mathcal{D}_{\textbf{W}} \log \tr[e^{-\beta h_{\textbf{W}_1}(\lambda)}] +\mathcal{D}_{\textbf{W}} \log \tr[e^{-\beta h_{\textbf{W}_2}(\lambda)}] \nonumber \\ &=0.
\end{align}
This means we can write the moments in terms of connected clusters only
\begin{equation}
\beta^m K_m = \sum_{\textbf{W} \in \mathcal{G}_m} \mathcal{D}_{\textbf{W}} \log \tr[e^{-\beta H(\lambda)}].
\end{equation}

%To relate this back to the original moments, we simply need to set all of the $\lambda_i=1$, as $K_m(1)=K_m$.

 This reduces the number of contributions to $K_m$ dramatically, and makes it possible to estimate them. One way to show the convergence of the series (see \cite{Kuwahara_2020_Gaussian,haah2021optimal,Wild_2023}) is to prove the following:
\begin{itemize}
    \item The number of connected clusters of size $m$ is bounded by $ N c_1^m$ for some constant $c_1$ \cite{Malyshev_1980,dobrushin1996estimates}.
    \item The size of each cluster derivative for a cluster of size $\vert \textbf{W}\vert=m$ is at most
    \begin{equation}
        \left \vert \mathcal{D}_{\textbf{W}} \log \tr[e^{-\beta H(\lambda)}] \right \vert \le (\beta c_2)^m m!
    \end{equation}
    for some constant $c_2$ \cite{haah2021optimal,Wild_2023}. 
\end{itemize}
The constants here can usually be taken to be simple functions of the lattice parameters $ \frak{d},k,J,h$ and of some property of the interaction graph. \new{For instance, in \cite{haah2021optimal,Wild_2023}, it is shown that $c_1=e \frak{d}$, and that $c_2=2eh(\frak{d}+1)$ (see Eq. \eqref{eq:h} for the definition of $h$). These facts together imply that $\vert K_m \vert \le 2e^2h \frak{d}(\frak{d}+1) \equiv (\beta^*)^{-1} \simeq \mathcal{O}(h)$, so that the partition function is analytic within a disk in the complex plane of radius $\beta < \beta^*$, and is also well approximated by its Taylor series. }

\new{Beyond this argument for the convergence of the series, there are other more general abstract methods for proving convergence of this type of quantity, in terms of the so-called \emph{polymer models} \cite{kotecky1986cluster,dobrushin1996estimates,Fernandez_2007,Mann2021}. See Chapter 5 of 
\cite{friedli2017} for an introduction.}

So far we have only discussed convergence of the series. However, the cluster expansion can be used to device efficient approximation schemes. The main idea is to prove that the individual Taylor terms can be computed efficiently. This requires two separate steps:
\begin{itemize}
 \item The set of all clusters of size $m$ can be enumerated in time $\text{poly}(N) \times \exp(\mathcal{O}(m))$ \cite{helmuth2020algorithmic,haah2021optimal}
 \item Each cluster derivative can be computed exactly in time $\text{poly}(N) \times \exp(\mathcal{O}(m))$ \cite{ComputingTutte,Mann2021,haah2021optimal}.
 \end{itemize}
We can thus add all the contributions from the different derivatives to obtain $K_m$ in time $\text{poly}(N) \times \exp(\mathcal{O}(m))$. This, together with Eq. \eqref{eq:clusterconvergence}, implies that by calculating the Taylor series up to a degree $M= \mathcal{O}\left(\log(N/\epsilon) \times \log(\beta^*/\beta)\right )$ there exists an $\epsilon$-close additive approximation to $\log Z$ that can be computed in time $\text{poly}(N, \epsilon^{-1})$. 

 We do not expect to be able to prove many general statements at all temperatures, due to the presence of thermal phase transitions, and to the fact that the ground state energy is computationally hard to estimate \cite{QMA2005}. However, there are specific models in the literature for which the convergence can be guaranteed for larger ranges of temperatures (see e.g. \cite{helmuth2022efficient} and references therein). See Sec. 
\ref{sec:Qpartition} for more details.

 This same technique also allows for e.g. the computation of expectation values such as $\tr[h_i e^{-\beta H}/Z]$ by differentiating by an extra $\lambda_i$ in the cluster derivative. It can be also applied to other similar objects such as characteristic functions of the form $\tr[e^{\alpha A} e^{-\beta H}/Z]$ for some local observable $A$, which allows for the derivation of probability theory statements, as explained further in Sec. \ref{sec:concentration}.

\subsection{Thermal locality estimates}\label{sec:localityestimates}

We now show the first method to decompose the thermal state into a product of smaller operators despite the non-commutativity, which is related to the general idea of operator growth. Consider an operator $A$ with local support on some small region on the lattice. For simplicity, let this region be such that $\vert \supp{A}\vert \le k$. 

An interesting quantity to study is the operator evolved in Euclidean or imaginary time $\beta$ under the Hamiltonian $H$,
\begin{equation}
  A(i\beta)=  e^{-\beta H} A e^{\beta H}.
\end{equation}
This is in analogy with the Heisenberg-picture operator $ A(t)= e^{i t H} A e^{-i t H}$, which can be understood in terms of the well-known Lieb-Robinson bounds \cite{Lieb2004}, that state that the support of $A(t)$ is mostly confined to a linear lightcone. In many situations, one will want to choose $A$ here to be one of the $h_i$ operators. 

It then makes sense to ask the following question: what is the locality of the Euclidean-evolved operator $A(i\beta)$? Perhaps surprisingly, this can be dramatically different to the real-time case: there is no general linear growth with the inverse temperature $\beta$, but a much wilder dependence on it. 

The main difference is that $e^{-i t H}$ is a unitary matrix, while $e^{-\beta H}$ is not. This means that results that exploit unitarity, such as the aforementioned Lieb-Robinson bounds, do not apply straightforwardly. Our best way forward seems then to analyze $A(i\beta)$ in terms of nested commutators
\begin{align}
    A(i\beta)&=\sum_{m=0}^\infty \frac{(-\beta)^m}{m!}[H,[H,....,[H,A]...] \nonumber \\ & \equiv \sum_{m=0}^{\infty} \beta^m C_m(A)  . \label{eq:nested}
\end{align}
It is easy to see that the $m$-th term in this expansion has support on a connected region whose furthermost point is a distance $m$ away from $A$. The question then becomes: how does this expansion in terms of $\beta$ converge? \new{We now discuss how the answer to this question (either high temperatures and 1D) appears to also be restricted to the one-phase region of completely analytical interactions from Sec. \ref{sec:completely}.}

It can be shown that, for general interaction graphs,  \cite{ruelle1999statistical,kuwahara2016}
\begin{equation}\label{eq:mthcoeff}
    \vert \vert C_m(A) \vert \vert \le k \vert \vert A \vert   \vert  (2 J k)^m,
\end{equation}
\new{with $J,k$ as defined in Sec. \ref{sec:latticenotation}.}
This statement is very much related to the bound on the number of connected clusters in Sec. \ref{sec:connectedclusters} above, since only connected clusters contribute to the nested commutators. Eq. \eqref{eq:mthcoeff} implies that as long as $\beta< (2J k) $, the expansion can be controlled as a geometric series, from which we obtain
\begin{align} \label{eq:normAbeta}
   & \vert \vert A(i\beta) \vert \vert \le k \vert \vert A \vert \vert \frac{1}{1- 2 \beta Jk} \\
   & \vert \vert A(i\beta) - \sum_{m=0}^{M} \beta^m C_m(A) \vert \vert \le  k \vert \vert A \vert \vert \frac{(2 \beta Jk)^{M+1}}{1- 2 \beta Jk}.\label{eq:normAbeta2}
\end{align}
Given that the $m$-th nested commutator can have support on at most $k \times m $ sites, the latter equation means that $A(i\beta)$ is, roughly speaking, localized within the subset of vertices a distance at most $k \times m$ away from $\supp{A}$. 
It is known that one cannot extend this result to temperatures lower than $\beta \simeq \mathcal{O}(1)$, since there exists a 2D lattice in which the terms in the nested commutators in Eq. \eqref{eq:nested} add up constructively, in a way that the norm of $ A(i\beta)$ \new{grows with system size, and diverges as $N \rightarrow \infty$} \cite{bouch2015}.

On the other hand, it has been known for some time \cite{araki1969} that when the lattice is a one-dimensional chain, the nested commutators grow more slowly, so that this type of convergence happens for all temperatures. For simplicity, we show explicitly the result for $k=2$ combining \cite{perezgarcia2020locality} and \cite{bouch2015}, which is
\begin{align}\label{eq:bound1D}
   & \vert \vert A(i\beta) \vert \vert \le  \vert \vert A \vert \vert f(\beta,J) \exp( f(\beta,J))\\
   & \vert \vert A(i\beta) - \sum_{m=0}^{M} \beta^m C_m(A) \vert \vert \le  15 \vert \vert A \vert \vert e^{-(M+1)} \nonumber\\ &\quad \quad \quad \forall M > g(\beta,J). \label{eq:bound1D2}
\end{align}
Here, we have defined $f(\beta,J) \equiv 16 \beta J \exp(1+8 \beta J)$ and $g(\beta,J) \equiv \exp(240 e^2 \beta J )-1$ \footnote{The numerical constants in these functions are likely sub-optimal, but this is very rarely important for applications.}. The intuitive reason for these is that the geometric bound of Eq. \eqref{eq:mthcoeff} can be improved in this case as \cite{bouch2015} (again, for $k=2$)
\begin{equation}
    \vert \vert C_m(A) \vert \vert \le 15 \vert \vert A \vert \vert \left( \frac{240 e J }{\log(m+1)}\right)^m.
\end{equation}
Notice that, because of the logarithm, the series in Eq. \eqref{eq:nested} is not geometric, and converges for all $\beta$. For further explanations of these points see also \cite{Avdoshkin2020}.

So far, we have described how does $A(i \beta)$ approximate its Taylor expansion. An alternative approximation commonly considered is to the operator $e^{-\beta H_{\Lambda_m}} A e^{\beta H_{\Lambda_m}}$, where $H_{\Lambda_m}= \sum_{\supp{h_i} \in \Lambda_m} h_i$ and $\Lambda_m$ \new{is a subset of the hypergraph corresponding to some region $V_m$ of the full lattice $\Lambda$}. One can then consider how the norm
\begin{equation} \label{eq:localHev}
    \vert \vert A(i \beta)- e^{-\beta H_{\Lambda_m}} A e^{\beta H_{\Lambda_m}} \vert \vert
\end{equation}
decays with $m$ in terms of how $\Lambda_m$ is defined (typically, some hyper-sphere centered around $A$). The analysis and convergence turn out to be almost the same as the one for the moments $C_m(A)$ above. The reason is that the difference between  $\sum_{m=0}^{M} \beta^m C_m(A)$ and $ e^{-\beta H_{\Lambda_m}} A e^{\beta H_{\Lambda_m}}$ are essentially the higher order terms in $\beta$ of the latter, which are also suppressed. See e.g. \cite{perezgarcia2020locality} for a detailed analysis of the 1D case or e.g. Lemma 20 in \cite{kuwahara2020} for a proof in higher dimensions.

One of the main reasons why both of these approximations are interesting is that they are related to the following propagator 
\begin{equation}\label{eq:transferop}
  E_A\equiv  e^{-\beta(H+A)}e^{\beta H} = \mathcal{T} \left( e^{-\int_0^\beta A(s) \text{d}s } \right),
\end{equation}
where $A(s)=e^{-s H} A e^{s H}$ and $\mathcal{T}$ denotes the usual time-ordered integral. This is such that
\begin{equation}\label{eq:leftmult}
    e^{-\beta (H+A)}= E_A e^{-\beta H}.
\end{equation}
\new{This operator $E_A$ can be used, for instance, to decompose $e^{-\beta H}$ as a product of its parts by e.g. choosing $A$ as the Hamiltonian at the boundary of two regions.}
This operator can be analyzed through a usual Dyson series in terms of powers of $e^{-x H} A e^{x H}$. \new{Assuming $\beta < (2Jk)^{-1}$}, it can be shown that $E_A$ has bounded norm as it follows from Eq. \eqref{eq:normAbeta} and \eqref{eq:transferop} that
\begin{equation}\label{eq:normEa}
    \vert \vert E_A \vert \vert \le \exp \left(\int_0^\beta \text{d}s \vert \vert e^{-s H} A e^{s H}\vert \vert  \right) \le \left( \frac{1}{1- 2 \beta Jk}  \right)^{\frac{ \vert \vert A \vert \vert}{2 \beta J}}.
\end{equation}
In Appendix \ref{app:localproof} we also show that it is approximately localized in a similar way as $e^{-x H} A e^{x H}$ is. This means that there exist an operator $E_A(l)$ with support restricted to a distance at most $l$ away from $A$ such that for $\beta < (2Jk)^{-1}$
\begin{equation}
       \vert \vert E_A -  E_A(l) \vert \vert \le \beta k \vert \vert A \vert \vert  \frac{(2 \beta Jk)^{l+1}}{(1- 2 \beta Jk)^{\frac{\norm{A}}{2\beta J}+1}}.
\end{equation}
 Also, notice that if $[H,A]=0$, then $ E_A= e^{-\beta A}$. With the right choice of $H,A$, the operator $E_A$ can thus be thought of as a ``transfer operator". \new{Corresponding results also exists for 1D using Eq. \eqref{eq:bound1D} and \eqref{eq:bound1D2}.}

Alternatively, one can also define the following operator
\begin{equation}
      E'_A\equiv  e^{-\beta(H+A)}e^{\beta H} e^{\beta A},
\end{equation}
with the difference that $H$ and $A$ are now treated on equal footing.
In this case, $E'_A$ is just the multiplicative error term in the first order Trotter product formula, which can be similarly analyzed through the expansion of $A(i \beta)$ (see the thorough analysis of Trotter errors in \cite{Childs_2021} for more details). These Trotter errors are most commonly analyzed in the context of digital quantum simulation \cite{Lloyd1996}, for which it is often convenient to go to higher orders in the decomposition. 

We finish this subsection outlining a result in 1D related to this discussion, which follows from bounds on the quantity in Eq. \eqref{eq:localHev}. It appeared first in \cite{araki1969}, and it features in Sections \ref{sec:local1D} and \ref{sec:algo1D}. Let us define $E^l_A=e^{-\beta(H_l+A)}e^{\beta H_l}$, where $H_l$ are the interaction terms a distance at most $l$ away from $\supp{A}$. It can be shown that 
\begin{align} \label{eq:araki1}
\vert \vert E_A \vert \vert &\le C_1 
\\ \label{eq:araki2}
\vert \vert E_A- E^l_A \vert \vert &\le C_2 \frac{q^l}{(l+1)!} ,
\end{align}
where $C_1,C_2$ and $q>1$ are constants depending on $k, J,\beta$ which we do not show explicitly for simplicity, although notice that $C_1$ will be essentially the exponential of Eq. \eqref{eq:bound1D} . The proof is similar to that of Appendix \ref{app:localproof}, together with a bound on Eq. \eqref{eq:localHev}. We refer the reader to e.g. \cite{araki1969,perezgarcia2020locality} for further details.

\new{The approximations $E_A(l)$ and $E_A^l$ to the operator $E_A$ are important in that they allow us to decompose $e^{-\beta H}$ into a product of smaller local operators despite the Hamiltonian being non-commuting. For instance, they will be useful in the arguments of Sec. \ref{sec:local1D}.}

\subsection{Quantum belief propagation}\label{sec:QBP}

An idea related to the previous locality estimates appeared first \cite{Hastings_2007}, and has more recently featured in several results about Gibbs states on lattices \cite{Kim2012,Kato_2019,EjimaQBP2019,brandao2019finite,Harrow_2020,kuwahara2020,Anshu_2021}.
\new{It is a tool similar to that of the previous Sec. \ref{sec:localityestimates}, in that it also allows us to decompose the thermal state as a product of smaller, localized operators, which makes certain calculations more tractable. The goal is to be able to divide the big operator $e^{-\beta H}$ into smaller pieces, that allow, for instance, to prove that the Gibbs state can be sequentially generated, or that local perturbations only have effect in the near vicinity.} 

This is part of a celebrated series of works including the decay of correlations for gapped ground states \cite{HastingsKoma2006} or the area law of entanglement in one dimension \cite{Hastings2007} which show how Lieb-Robinson bounds (a dynamical statement) can be used to prove static properties about ground and thermal states.\new{ The derivation here is a particular example of that idea, but see \cite{hastings2010locality,HastingsICM2021} for overviews that go beyond Gibbs states.} 

\new{The goal is to construct a quasi-local operator $O_A^m$ (to be defined below) with support near $\text{supp}(A)$  such that
\begin{equation}
    e^{-\beta (H+A)} \simeq O_A^m e^{-\beta H}(O_A^m)^\dagger.
\end{equation}
Notice the difference with Eq. \eqref{eq:leftmult}, where we only multiply $e^{-\beta H}$ with an operator from the left.}

We start by considering the ``perturbed" Hamiltonian $H(s)=H+sA$ and the following derivative
\begin{equation}\label{eq:derivativeQBP}
    \frac{d e^{-\beta H(s)}}{d s}=-\frac{\beta}{2} \left \{ e^{-\beta H(s)}, \Phi_\beta^{H(s)}(A) \right \}
\end{equation}
where, if $H(s)=\sum_i E_i (s) \vert{i(s)}\rangle \langle{i(s)}\vert$ is the energy eigenbasis,
\begin{equation}
   \Phi_\beta^{H(s)}(A)_{ij}= \langle i(s) \vert A \vert j(s) \rangle \tilde f_\beta(E_i(s)-E_j(s)),
\end{equation}
where $\tilde f_\beta (\omega)=\frac{\tanh(\beta \omega /2)}{\beta \omega/2}$. With $f_\beta(t)=\frac{4}{\beta \pi}\log \left( \frac{e^{\pi \vert t \vert}{\beta}+1}{e^{\pi \vert t \vert}{\beta}-1} \right)$ the Fourier transform of $\tilde f_\beta (\omega)$ (see Appendix B of \cite{Anshu_2021}), we can also write
\begin{equation}\label{eq:beliefprop1}
    \Phi_\beta^{H(s)}(A) = \int_{-\infty}^\infty \text{d}t f_\beta (t) e^{-i tH(s)} A e^{i tH(s)}.
\end{equation}
\new{The proof leading to Eq. \eqref{eq:derivativeQBP} that explains hte appearance of $\tilde f_\beta (\omega)$ is shown in App. \ref{app:QBP}.}

\new{Since $\vert \vert e^{-i tH(s)} A e^{i tH(s)} \vert \vert = \norm{A}$, $\norm{\Phi_\beta^{H(s)}(A)} \le \vert \vert A \vert \vert $ by the triangle inequality and Eq. \eqref{eq:normft}. Moreover, it can also be approximated by a localized operator around the support of $A$ by using Lieb-Robinson bounds \cite{Lieb2004}. In particular, when $H(s)$ is local, it can be shown that \cite{Barthel_2012}
\begin{align}
    \vert \vert e^{-i tH(s)} A e^{i tH(s)} &- e^{-i t H_{\Lambda_m}(s)} A e^{i t H_{\Lambda_m}(s)} \vert \vert \nonumber \\ &\le  m^{D-1} b \vert \vert A \vert \vert e^{c' (vt-m)},\label{eq:LRB}
\end{align}
where $H_{\Lambda_m}(s)$ is the restriction of $H(s)$ to the sum of local terms that are at most a distance $m$ away from the support of $A$, $v$, $b$ and $c'$ are constants, and $D$ is the dimension of the interaction lattice. This should be reminiscent of the  $H_{\Lambda_m}$ appearing in Eq. \eqref{eq:localHev}, with the only difference being that now the evolution is for real times. }

\new{
Eq. \eqref{eq:LRB} allows us to establish that after time-evolving $A$ for a short time, the support is still approximately localized, with a radius growing with time.  This also allows us to define $ \Phi_\beta^{H_{\Lambda_m}(s)}(A)$, which is close to the original $ \Phi_\beta^{H(s)}(A)$ in the following sense
\begin{align} \label{eq:QBPDist}
&\frac{1}{\vert \vert A \vert \vert}\vert \vert  \Phi_\beta^{H(s)}(A)  -   \Phi_\beta^{H_{\Lambda_m}(s)}(A) \vert \vert \\ \nonumber &\le  m^{D-1} b   \int_{-m/2v}^{m/2v} e^{c' (v\vert t \vert-m)} f_\beta(t) \text{d}t + 4  \int_{m/2v}^{\infty} f_\beta(t) \text{d}t   \\ \nonumber & \le m^{D-1} b e^{-c' m/2} +  \frac{1}{ e^{\frac{\pi m}{2\beta v}}-1},
\end{align}
where in the first line, after the triangle inequality, we divided the integral into two different ranges, and used Eq. \eqref{eq:LRB} in the first range and $\norm{\Phi_\beta^{H(s)}(A)} \le \vert \vert A \vert \vert $ in the second. The bounds on the integrals were obtained using the properties of $f_\beta(t)$ from Eq. \eqref{eq:normft} and \eqref{eq:ftdecays}.
As a result, for $m$ large enough, the difference between the two operators is exponentially decaying in $m$.
}

We can now integrate Eq. \eqref{eq:derivativeQBP} between $s=0$ and $s=1$ to obtain
\begin{equation}
    e^{-\beta (H+A)}= O_A e^{-\beta H} O_A^\dagger,
\end{equation}
where 
\begin{equation}\label{eq:QBPoperator}
    O_A = \mathcal{T} e^{-\frac{\beta}{2}\int_0^1 \text{d}s \Phi_\beta^{H(s)}(A)  }.
\end{equation}
\new{Similarly to Eq. \eqref{eq:transferop} above (see the analyisis of the operator $E_A$ in  Appendix \ref{app:localproof}), this operator has a bounded norm, since
\begin{align}\label{eq:QBPnorm}
    \vert \vert O_A \vert \vert \le e^{\frac{\beta}{2}\int_0^1 \text{d}s \vert \vert \Phi_\beta^{H(s)}(A) \vert \vert  }  = e^{\frac{\beta}{2} \vert \vert A \vert \vert},
\end{align}
Additionally, it is also approximately localized exactly around the support of $A$. Let us define the operator $O_A^{m}$ in the natural way
\begin{equation}\label{eq:QBPoperator2}
    O_A^m \equiv \mathcal{T} e^{-\frac{\beta}{2}\int_0^1 \text{d}s \Phi_\beta^{H_{\Lambda_m}(s)}(A)  }.
\end{equation}
We can use an argument analogous to that used in App. \ref{app:localproof} to show that $O_A$ and $O_A^m$ are exponentially close, as, given Eq. \eqref{eq:QBPDist}, for large enough $m$,
\begin{align}\label{eq:QBPlocal}
    \vert \vert O_A - O_A^{m} \vert \vert & \le \frac{\beta \norm{O_A}}{2} \int_0^1 \text{d}s \vert \vert  \Phi_\beta^{H(s)}(A)  -   \Phi_\beta^{H_{\Lambda_m}(s)}(A) \vert \vert \\ & \le  \beta  \norm{A} e^{\frac{\beta \norm{A}}{2}-\Omega(m)}.
\end{align}}

%\alv{old part}
%This is such that
%\begin{align}\label{eq:QBPlocal}
%    \vert \vert O_A - O_A^{l} \vert \vert \le  \frac{c' \beta \norm{A}}{2} e^{(1+c')\frac{\beta}{2} \vert \vert A \vert \vert}e^{-\frac{c'l}{1+c' v \frac{\beta}{\pi}}}.
%\end{align}

These bounds can be compared to Eq. \eqref{eq:araki1} and \eqref{eq:araki2}, which are of a very similar nature. There are, however, two important differences between $O_A$ and $E_A$ in Eq. \eqref{eq:transferop} above: 
\begin{itemize}
    \item Since it is based on the Lieb-Robinson bound, the operator $O_A$ is well-behaved in all lattices and at all temperatures, in the sense that it has a bounded norm and is approximately localized. This is as opposed to $E_A$, which is likely a large operator in high dimensions and low temperatures. \new{That is, this result holds for all Gibbs states, irrespective of whether they are completely analytical interactions or not.}
    \item On the other hand, to recover $e^{-\beta (H+A)}$ from $e^{-\beta H}$ we require left and right multiplication with $O_A,O_A^\dagger$, as opposed to Eq. \eqref{eq:leftmult}, which may be problematic in some applications. In particular, we should not expect it to be a key ingredient in proving results that do not hold at all temperatures such as the decay of correlations or the analyticity of the partition function.
\end{itemize}

\subsection{Selected trace inequalities}\label{sec:traceineq}

In the past two subsections we have explored how to analyze perturbations to the Hamiltonian in the Gibbs operator $e^{-\beta (H+A)}$ via $E_A$ and $O_A$. When considering traces, simpler identities hold. We exemplify this with two with very elementary implications and proofs, which can be found in (at least) \cite{Lenci2005}. The first one is about the stability of partition functions. Let $H_1,H_2$ be Hermitian operators. Then we have
\begin{equation}\label{eq:mineq1}
    \left \vert \log \tr[e^{H_1+H_2}]- \log \tr[e^{H_1}] \right \vert \le \norm{H_2}.
\end{equation}
The proof is just as follows:
\begin{align}
    &\left \vert  \log \tr[e^{H_1+H_2}]- \log \tr[e^{H_1}] \right  \vert \nonumber \\& \quad= \left \vert \int_0^1 \frac{\text{d}}{\text{d}t}\log \tr \left[ e^{H_1 + t H_2} \right ]  \text{d}t \right \vert \nonumber \\& \quad \le  \int_0^1 \left \vert \frac{\tr[H_2e^{H_1 + t H_2} ]}{\tr[e^{H_1 + t H_2}]} \right \vert \text{d}t \nonumber \\& \quad \le \vert \vert H_2 \vert \vert,
\end{align}
where in the last inequality we have simply used H\"older's inequality Eq. \eqref{eq:Holder} with $q_1=1,q_2=\infty$. 
If we take e.g. $H_1=-\beta H,H_2=-\beta A$, this implies that changing the Hamiltonian by $A$ changes the log-partition function at most by $\beta \norm{A}$.

The second is a similar result that holds for expectation values of positive operators. Let $H_1,H_2$ be as before, and let $C>0$. Then 
\begin{align}\label{eq:mineq2}
  &    \left \vert \log \tr[C e^{H_1+H_2}]- \log \tr[C e^{H_1}] \right \vert \\ \nonumber &\quad \le \int_{0}^1 \text{d}t \int_{-1/2}^{1/2} \text{d}s \norm{e^{-s(H_1+tH_2)}H_2 e^{s(H_1+tH_2)}}.
\end{align}
The proof can be found in Appendix \ref{app:mineq2}. This norm can then be bounded with the results from Sec. \ref{sec:localityestimates}, to scale as $\propto \vert \vert H_2 \vert \vert$. The resulting expression can for instance be used for analyzing characteristic functions of observables $F$ by taking $C=e^{\alpha F}$ for $\alpha \in \mathbb{R}$ \cite{Lenci2005}.

More generally, in the practice of mathematical quantum physics, whether it is from the many body, the QI, or any other perspective, many important proof ingredients take the form of inequalities, either between operators, traces, or norms (such as those already mentioned in Sec. \ref{sec:norms}). There are too many to give a reasonably complete overview here but we refer the reader to e.g. \cite{bhatia2013matrix,carlen2010trace}.

%%%%%%%%%%%
%%%%%%%%%%%

\section{Correlations}\label{sec:correlations}

One of the more important questions when studying many body systems is: how and how much are the different parts correlated? Intuitively, the stronger these correlations, and the longer their range, the more complex a state is - the reason being that we cannot think of the large system as a collection of simpler, weakly correlated parts. The obvious extreme example is that of an uncorrelated gas, in which the particles do not interact and have \new{completely independent properties}.

For thermal states with local interactions, we can expect that locality will make the state far from generic, in a way that constraints its complexity. Intuitively, it should cause the correlations to be ``localized", meaning that particles are only correlated with their vicinity as given by the lattice geometry. For a rough intuition, consider the first terms of the Taylor series
\begin{equation}
    e^{-\beta H} = \mathbb{I} - \beta \sum_{i} h_i +\frac{\beta^2}{2}\sum_{i,j} h_i h_j +...
\end{equation}
That is, at very high temperatures we approach the trivial uncorrelated state $\propto \mathbb{I}$ and the leading order term includes only $k$-local couplings, with only higher order terms coupling far away particles. We thus expect that the correlations between particles will generally be weaker $i)$ the higher the temperature and $ii)$ the larger their distance on the interaction graph. \new{This is one of the main ways of understanding how the Gibbs states of the ``one phase region" of Sec. \ref{sec:completely} are very different from generic states.}

An important motivation for this is that, as we will see in later sections, the situations in which the correlations are weaker or short range roughly correspond to those in which we expect better algorithms for the description of thermal states. This is perhaps most clearly the case in the context of tensor network methods. We now proceed to describe (and even prove) the more important ways in which these correlations are constrained.

\subsection{Correlations between neighbouring regions: Thermal area law}\label{sec:arealaw}

One of the more important statements about correlations in quantum many-body systems is the area law. This roughly states that a measure of correlations between two adjacent regions is upper-bounded by a number proportional to the size of their mutual boundary. 

Traditionally, this has been mostly studied in the context of ground states, which are pure. There, the \new{relevant} measure of correlations is the entanglement entropy, or some R\'enyi version of it. In that context, an area law for the entanglement entropy is believed to hold for all ground states of models with a gap \cite{Eisert2010}. This can be proven in 1D \cite{Hastings2007,arad2013area} and in some cases in 2D \cite{anshu2021area}. The interest in it is largely due to its relation to other phenomena, such as phase transitions \cite{Vidal2003}, the decay of long-range correlations \cite{Brandao2014} or the effectiveness of certain tensor network algorithms \cite{verstraete2006matrix,Ge_2016_2}. 

For thermal states, a very general area law can be shown to hold for systems in any dimension, at all temperatures. We now give a short proof of this statement, and then discuss its significance (see \cite{wolf2008} for the original reference).
In this case, since it is a mixed state, an appropriate measure of correlations is the mutual information in Eq. \eqref{eq:MI}. 

Let us partition our interaction graph into two subsets of particles $A,B$, with a thermal state $\rho_\beta^{AB}$. We start with the very simple thermodynamic observation that the free energy $F$ \new{from Eq. \eqref{eq:relFree}} of the thermal state is lower than that of any other state (this follows from Eq. \eqref{eq:relFree}), and in particular 
\begin{equation}
    F_\beta(\rho_\beta^{AB}) \le F_\beta(\rho_\beta^{A} \otimes \rho_\beta^{B} ).
\end{equation}
Writing out the free energy explicitly as $F_\beta(\rho)= \tr[\rho H ] - \beta^{-1} S(\rho)$ and rearranging yields
\begin{equation}\label{eq:miE}
S(\rho_\beta^{A} \otimes \rho_\beta^{B})- S(\rho_\beta^{AB}) \le \beta \left( \tr[H \rho_\beta^{A} \otimes \rho_\beta^{B}]-\tr[H \rho_\beta^{AB}] \right).
\end{equation}
Given that the entropy is additive $S(\rho \otimes \sigma)=S(\rho)+S(\sigma)$ notice that the LHS is exactly the mutual information $I(A:B)_{\rho_\beta^{AB}}$ from Eq. \eqref{eq:MI}. Since the Hamiltonian is local, we can write it as
\begin{equation}
    H=H_A+H_B+H_I,
\end{equation}
where $H_A,H_B$ have support only on $A,B$ respectively, and $H_I$ is the interaction between them (with support on both). By definition, the expectation values of $H_A$ and $H_B$ coincide on both states $\tr[(H_A+H_B ) \rho_\beta^{A} \otimes \rho_\beta^{B}]=\tr[(H_A+H_B ) \rho_\beta^{AB}]$, so that 
\begin{align}
    & \beta \left( \tr[H \rho_\beta^{A} \otimes \rho_\beta^{B}]-\tr[H \rho_\beta^{AB}] \right) \\ \nonumber &= \beta \left( \tr[H_I \rho_\beta^{A} \otimes \rho_\beta^{B}]-\tr[H_I \rho_\beta^{AB}] \right). 
\end{align}
Now we can use a few of the operator inequalities from Section \ref{sec:norms} to obtain
\begin{align}\label{eq:enB}
    \tr[H_I \rho_\beta^{A}& \otimes \rho_\beta^{B}]-\tr[H_I \rho_\beta^{AB}]  \le \vert \vert H_I (\rho_\beta^{A} \otimes \rho_\beta^{B}-\rho_\beta^{AB}) \vert \vert_1
    \nonumber\\ & \le \vert \vert H_I \vert \vert \times  \vert \vert \rho_\beta^{A} \otimes \rho_\beta^{B}-\rho_\beta^{AB} \vert \vert_1
    \\ \nonumber &\le \vert \vert H_I \vert \vert \times ( \vert \vert \rho_\beta^{A} \otimes \rho_\beta^{B} \vert \vert_1+ \vert \vert \rho_\beta^{AB} \vert \vert_1 ) =2 \vert \vert H_I \vert \vert.
\end{align}
Putting Eq. \eqref{eq:miE} and \eqref{eq:enB} together we have the final result
\begin{equation}\label{eq:arealaw}
  I(A:B)_{\rho_\beta^{AB}}  \le 2 \beta \vert \vert H_I \vert \vert.
\end{equation}

\begin{figure}[t] 
\includegraphics[width=0.7\linewidth]{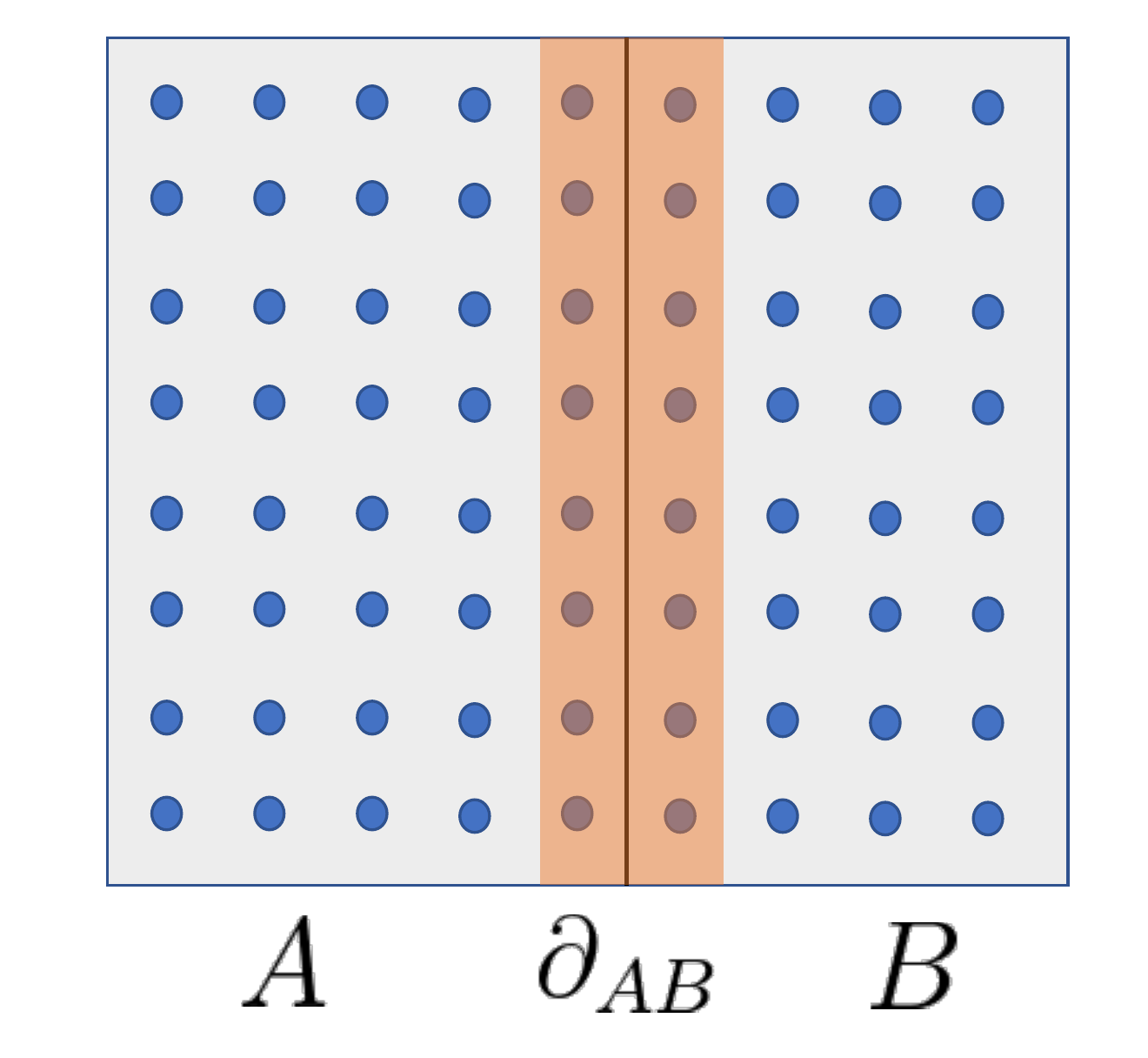}.
\caption{\new{Under an area law,} the correlations between regions A and B grow at most as \new{the size of their} mutual boundary $\partial_{AB}$.} \label{fig:arealaw}
\end{figure}

This is the \emph{area law} for the mutual information of a thermal state: it implies that the strength of the correlations of systems $A,B$ cannot depend on their size, but that it grows at most as their common boundary. For a local Hamiltonian, we have that
\begin{equation}
    \vert \vert H_I \vert \vert \leq 2 k h \vert  \partial_{AB}\vert  ,
\end{equation}
\new{where $\partial_{AB}=\partial_A \cup \partial_B$, $h$ is defined in Eq. \eqref{eq:h} and $k$ is the largest support of any $h_i$.} Notice that with $\vert  \partial_{AB}\vert$ we do not mean the size of the boundary of systems $A,B$ together, but the number of elements of $\partial_A$ that are connected to $\partial_B$ by hyperedges. We show this schematically in Fig. \ref{fig:arealaw}. This is to be contrasted with the most general upper bound on the mutual information, which is $I(A:B) \le \min \{\log(d_A),\log(d_B)\}$ (since $
\log d_A \propto \vert A \vert$ \new{the largest possible scaling is} a ``volume law" instead).

What this strongly suggests (although it does not quite prove) is that the correlations between $A$ and $B$ are localized around the mutual boundary, and that the bulks of $A$ and $B$ are mostly uncorrelated. That is, the only relevant information about $A$ that $B$ contains is about the region of $A$ that is near their boundary.

This statement, as can be seen from the proof, holds for all temperatures and all interaction graphs, which is likely as general as it can be\new{, at least for systems with finite-dimensional Hilbert spaces \cite{lemm2022thermal}}. The drawback of that generality, however, is that it will be unable to signal important phenomena that happens only at specific temperature ranges, such as thermal phase transitions, or an efficient classical or quantum simulability. \new{That is, Eq. \eqref{eq:arealaw} does not narrow down the set of ``completely analytical interactions" for Sec. \ref{sec:completely} in any meaningful way.}
Other more specific versions of the thermal area law in the literature may have more potential in this regard \cite{sherman2016,scalet2021computable}.

The temperature dependence of Eq. \eqref{eq:arealaw} can be improved to $\tilde {\mathcal{O}} (\beta^{2/3})$ \cite{kuwahara2020}. This can be proven with a variety of techniques, including those of Sec. \ref{sec:localityestimates} and Sec. \ref{sec:QBP}, as well as methods originated in the study of ground states \cite{arad2013area}. This dependence is not far from optimal, since there exists a 1D model for which the scaling of the MI is at least $\mathcal{O}(\beta^{1/5})$ at low temperatures \cite{Gottesman2010}. 

 Many important physical models have a very different temperature dependence, such as $\log(\beta+1)$ \cite{_nidari__2008,bernigau2015}. Classical systems, on the other hand, have an upper bound that is independent of the temperature, as $I(A:B)_{\rho_\beta^{AB}}  \le \vert \partial_{AB}\vert  \log d $ \cite{wolf2008}. All these suggests that the scaling of the mutual information with $\beta$ in the low temperature regime is related to the computational complexity of the ground space of the models.

\subsection{Decay of long-range correlations}\label{sec:corrdecay}

\begin{figure}[t]
\includegraphics[width=0.9\linewidth]{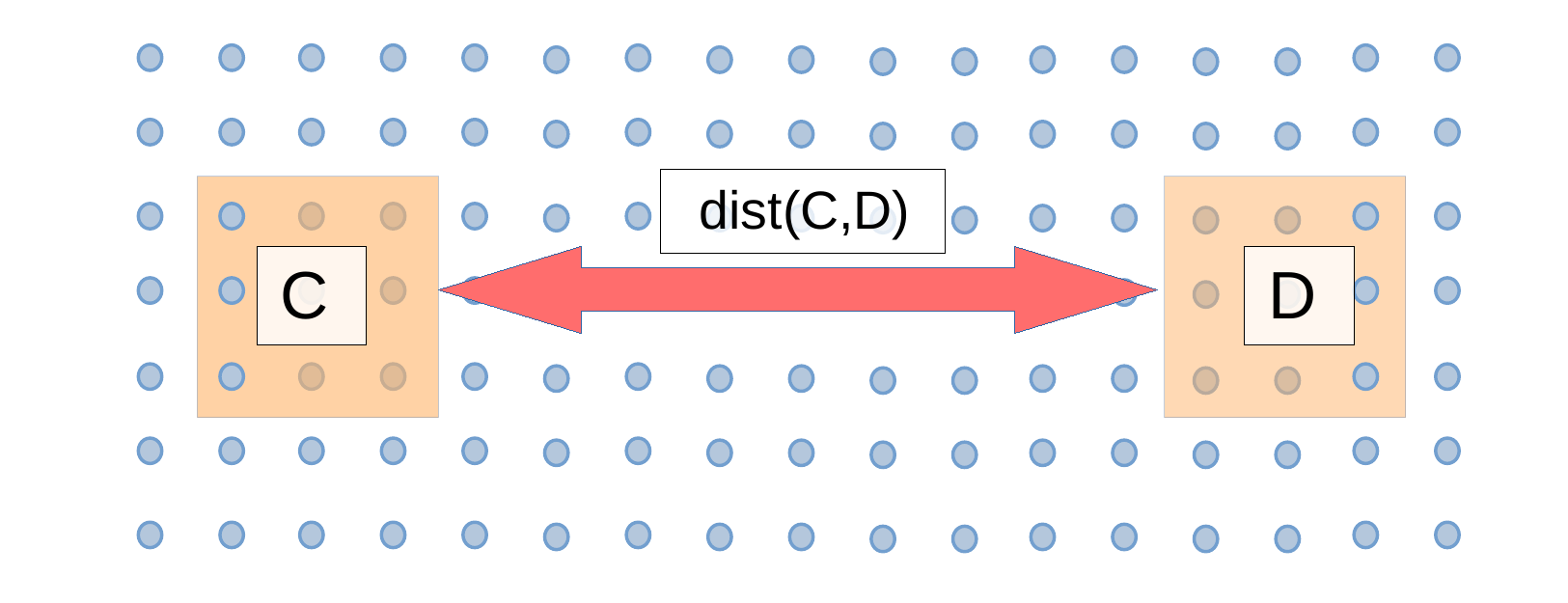}.
\caption{Regions $C,D$ in the lattice are separated by a distance $\text{dist}(C,D)$. The mutual information between these two regions typically decays exponentially with their distance.}\label{fig:distanceCD}
\end{figure}

An important fact about thermal states is that often their spatially separated parts are very weakly correlated. Let $C,D$ be regions such that their distance is $\text{dist}(C,D)$ (see Fig. \ref{fig:distanceCD}). We focus on measures of correlations evaluated at the marginals on these regions $\tr_{\setminus (CD)}[\rho_\beta]=\rho_\beta^{CD}$. For instance, taking the mutual information, we expect that in general
\begin{equation}
I(C:D)_{\rho_\beta^{CD}} \le f\left(\text{dist}(C,D)\right),
\end{equation}
where $f$ is some rapidly decaying function. In fact, \new{we expect that for completely analytical interactions} $f(l)\le K \vert \partial_C \vert \vert \partial_D \vert e^{-l/\xi}$, where $K>0$ is some constant, $\partial_{C,D}$ is the size of the boundary of each region, and $\xi$ is the thermal \emph{correlation length} that depends on the temperature and other parameters, but not on $l$ or the system's size. 

This has been proven in \new{translation invariant} 1D systems at all temperatures \cite{bluhm2021exponential}. The main idea behind it is to use the locality estimates from \ref{sec:localityestimates}, and in particular the properties of the operator $E_A$ in Eq. \ref{eq:leftmult}. \new{That this also holds for high enough temperatures in all dimensions follows the cluster expansion applied to the mutual information \cite{kuwahara2020}.}

%in the two following scenarios: \alv{we have to change this too!!}
%\begin{itemize}
%	\item For any $k$-local interaction graph above a threshold temperature $\beta < \beta^*$, where $\beta^*$ depends on parameters of the Hamiltonian (but not on its size) . This has been proven through the cluster expansion technique outlined in Sec. \ref{sec:connectedclusters} \cite{kliesch2014,Frohlich2015Some,}.
%	\item For 
%\end{itemize}
% The proofs are slightly involved and beyond the scope of this tutorial, so we refer the reader to the original references.

A more commonly stated but weaker condition is the decay of the connected two-point correlators. This usually takes the form
\begin{align}\nonumber
\frac{\left\vert \tr[\rho_\beta M_C \otimes M_D ]-\tr[\rho_\beta M_C] \tr[\rho_\beta M_D ] \right\vert }{\vert \vert M_C \vert \vert \vert \vert M_D \vert \vert} \\\le K  \vert \partial_C \vert \vert \partial_D \vert  e^{-\text{dist}(C,D) / \xi},\label{eq:corrdecay}
\end{align}
where here $M_C$ and $M_D$ have support on regions $C,D$, respectively. \new{That this is weaker than the decay of the mutual information follows from Pinsker's inequality applied to the marginal on regions $C,D$.}

\new{It is known that correlations decay exponentially at large temperatures for arbitrary interaction graphs \cite{kliesch2014,Frohlich2015Some}. This can be shown with the cluster expansion \cite{Ueltschi2004}. Instead of showing the proof in full generality, we can already see a simple but instructive case by noticing that, with the notation of Sec. \ref{sec:connectedclusters},
\begin{align}
    \frac{d \log Z (\lambda)}{ d \lambda_i d \lambda_j} \Big\vert_{\lambda=1} = \beta^2 (\langle h_i h_j \rangle - \langle h_i \rangle \langle h_j \rangle ).
\end{align}
At the same time, when one differentiates over two variables $\lambda_i,\lambda_j$, the nonzero contributions come from  clusters that contain both 
\begin{align}
& \frac{d \log Z (\lambda)}{ d \lambda_i d \lambda_j} \\&=  \frac{d }{ d \lambda_i d \lambda_j} \left(\sum_m \frac{1}{m!}
\sum_{\substack{\textbf{W} \in \mathcal{G}_m \\  i,j \in \textbf{W}}} \mathcal{D}_{\textbf{W}} \log \tr[e^{-\beta H(\lambda)}] \right).\nonumber
\end{align}
However, connected clusters such that $i,j \in \textbf{W}$ belong to $\mathcal{G}_m$ with $m \ge \text{dist}(i,j)$. This means that the lowest moment that appears in the correlation function is $K_m $ with $m=\text{dist}(i,j)$. If Eq. \ref{eq:Kmsmall} holds, then the correlation function will decay at least as  $(\beta/\beta^*)^{\text{dist}(i,j)}$, mirroring \eqref{eq:corrdecay}. A similar argument also holds for arbitrary few body-observables $h_C,h_D$ if one consider an appropriate cluster expansion of the perturbed Hamiltonian $H+h_C+h_D$. See e.g. \cite{Ueltschi2004,Frohlich2015Some} for more general results.}

\new{The connection between this type of correlation decay and the analiticity of the partition function is very well understood in the classical case, where they are known to be equivalent \cite{Dobrushin1987}.
In the quantum case, it is only known that a condition stronger than the analiticity of $\log Z$ (called \emph{analiticity after measurement}) implies decay of correlations. See \cite{Harrow_2020} for details.}

Intuitively, both these properties are related to the absence of phase thermal phase transitions: at those critical points, the correlation function diverges and the partition function becomes non-analytic. Since there are known phase transitions at finite temperature (e.g. 2D classical Ising model), the exponential decay does not hold for all thermal states at all temperatures. \new{As such, this can typically be thought of as a characteristic property of the set of completely analytical interactions from Sec. \ref{sec:completely}.}

The decay of correlations is an important fact: it shows that the different parts of the system behave almost completely independently. A state with this property should then share many large-scale features with an uncorrelated gas, in which the particles are not interacting at all. This has as a wealth of related physical consequences. For instance, it is associated with basic statistical physics facts covered in Sec. \ref{sec:statistical}, in particular the validity of the central limit theorem and related results on concentration properties of thermal states \cite{brandao2015equivalence,Anshu_2016} and the phenomenon of equivalence of ensembles \cite{brandao2015equivalence,Tasaki2018,Kuwahara_2020_ETH}. It also features in the proof of \emph{local indistinguishabiliy} in Sec. \ref{sec:local1D}.

\subsection{A refined correlation decay: Conditional mutual information}\label{sec:CMI}

In Sec. \ref{sec:arealaw}, we mentioned that the area law itself does not quite imply that the correlations in a system are localized, in the sense that a particular subsystem is only appreciably correlated with its vicinity. There is, however, a significantly stronger statement about correlations that does imply it in a clear way. \footnote{The ideas described here can be seen as the quantum analogues of a much stronger statement that holds for classical probability distributions: the Hammersley-Clifford theorem \cite{hammersley1971markov}.}

\begin{figure}[t]
\includegraphics[width=0.8\linewidth]{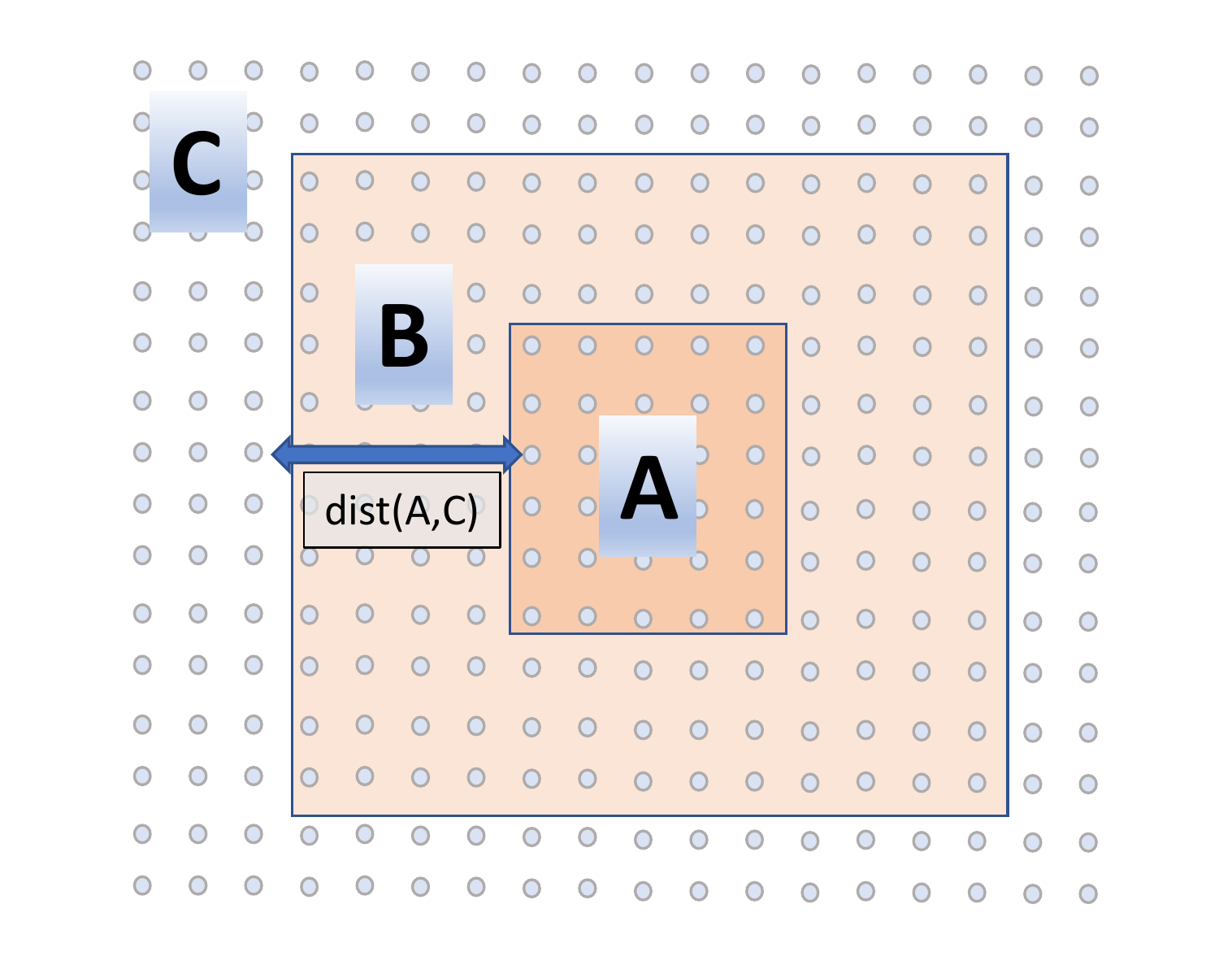}.
\caption{In this configuration, the region $B$ shields $A$ from $C$, such that the minimum distance between $A$ and $C$ is given by the shortest path from $A$ to $C$ through $B$.}\label{fig:shield}
\end{figure}

\begin{figure}[t]
\centering
\includegraphics[width=1\linewidth]{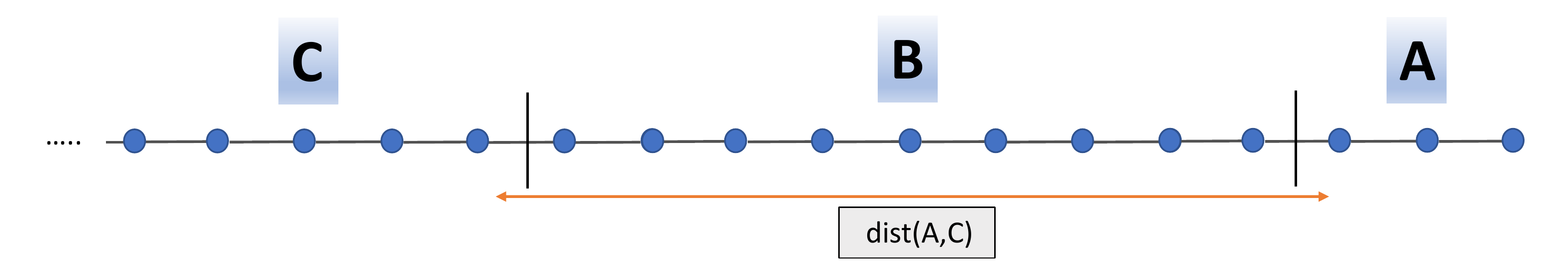}.
\caption{In this chain, the middle region $B$ shields $A$ from $C$, and their distance is related to the size of $B$.}\label{fig:shield1D}
\end{figure}

This is the property of being an approximate \emph{quantum Markov state} \cite{Hayden_2004}, which is defined in terms of the decay of the CMI in Eq. \eqref{eq:CMI}. Let us consider three regions $A,B,C$ such that $B$ shields $A$ from $C$. A simple example is given in Fig. \ref{fig:shield}, or in Fig. \ref{fig:shield1D} for 1D.

Since this quantifies how many of the correlations between $A$ and $C$ are not mediated through $B$,  we thus expect that it becomes small as the size of $B$ grows, and $A,C$ are further apart. This is perhaps the strongest sense in which correlations can be localized. 

\new{This is studied in one dimensional systems in \cite{Kato_2019}. By choosing $A,B,C$ to be adjacent regions of the chain (see Fig. \ref{fig:shield1D}), \cite{Kato_2019} shows that 
    \begin{equation}\label{eq:1DMarkov}
    I(A:C \vert B)_{\rho_\beta} \le c_1 \vert B \vert e^{- c_2\sqrt{\vert B \vert}}.
    \end{equation}
It is expected that the decay is $e^{- \Omega ( \vert B \vert)}$ as opposed to Eq. \eqref{eq:1DMarkov}, which may be important for certain applications  \cite{kim2017markovian,Kato_2019,brandao2019finite}.
The key technique is the quantum belief propagation from Sec. \ref{sec:QBP}, but the locality bounds from Sec. \ref{sec:localityestimates} are also sufficient. The idea is to use those results to define a completely positive map corresponding to a particular POVM outcome, that can be used in a ``measure until success" strategy. This result, however, relies on the exponential decay of correlations in 1D, and also on bound on the correlation length of the form $\xi \le e^{\mathcal{O}(\beta)}$, which is currently not known.  Interestingly, \cite{Kato_2019} also shows a converse statement: any state with a sufficiently fast decaying CMI approximates the thermal state of some local Hamiltonian.}
    
\new{In larger dimensions, the work \cite{Kuwahara_2020_Markov} shows that a non-commutative analogue of the cluster expansion (in which there are no traces or expectation values, but rather operators) suffices to study this problem. In particular, if the cluster expansion corresponding to the object $\log \tr_{\setminus A}[e^{-\beta H}]$ converges, exponentially, one obtains an exponential decay of the form
\begin{equation}
    I(A:C \vert B)_{\rho_\beta} \le k_1 \min \{ \vert \partial A \vert, \vert \partial C \vert \} \left(\frac{\beta}{\beta^*_{\text{NC}}}\right)^{- k_2 \times \text{dist}(A,C)},
    \end{equation}
    which only works for high enough temperatures $\beta < \beta^*_{\text{NC}}=\mathcal{O}(1)$.
    This expansion is more involved than the one described in Sec. \ref{sec:connectedclusters} in that the individual terms of the expansion may not commute (since no trace is being taken when expanding $\log \tr_{\setminus A}[e^{-\beta H}]$). Thus, the constant $\beta^*_{\text{NC}}$ need not be the same as the $\beta^*$.}

The significance of a fast decay of the CMI is highlighted by the idea of the Petz map \cite{petz1986sufficient,Hayden2004}. An important result in this regard states that, given a tripartite state $\rho=\rho^{ABC}$, there exists a CPTP map $\mathcal{N}(\cdot)_{B \rightarrow BC}$ (that is, acting on $B$, and with output on BC) such that \cite{Fawzi_2015,SutterRec2018} 
\begin{equation}\label{eq:recovery}
    I(A:C \vert B )_\rho \ge 2 \vert \vert \rho^{ABC} - \mathcal{N}(\rho^{AB})_{B \rightarrow BC} \vert \vert_1.
\end{equation}
 The map $\mathcal{N}$ usually goes under the name of \emph{recovery map}. See \cite{Sutter2018} for an overview.

A fast decay of the CMI thus guarantees that the Gibbs state on $A,B,C$ can be reconstructed from $\rho_{AB}$ by acting locally on $B$ (and importantly, not on $A$), such that $\mathcal{I}_A \otimes \mathcal{R}_{B\rightarrow BC} (\rho^{AB}) \simeq \rho^{ABC}$, with $\mathcal{R}_{B\rightarrow BC}$ some CP map taking only $B$ as input. This gives a way of sequentially preparing the whole thermal state from its smaller components, which can potentially be used e.g. for quantum algorithms (see Sec. \ref{sec:algorithms}). 

 %This says that a classical Gibbs state of a local Hamiltonian with interaction graph $\Lambda$ is also a so-called Markov random field defined in terms of the graph $\Lambda$. This means that for classical Hamiltonians, the CMI as defined here is always zero. In Sec. \ref{sec:commuting} we will see that a similar result also holds for \emph{commuting} quantum Hamiltonians.

%%%%%%%%
%%%%%%%%

\section{Locality of temperature}\label{sec:subsystem}

In the previous section we focused on the correlations between different parts. Now, we move the spotlight to features of individual subsystems. That is, if we divide the system into $A$ and its compliment $\setminus A$, what does $\tr_{\setminus A}[\rho_\beta]$ look like? In the rest of the section we drop the subscript $\beta$ for simplicity of notation.

Consider first the trivial case: if the particles are non-interacting, it holds that the marginal on $A$ is the thermal state of $H_A$ which is the Hamiltonian that acts only on subsystem $A$. That is
\begin{equation}\label{eq:nonin}
   \rho^A \equiv \tr_{\setminus A}[\rho]= \frac{e^{-\beta H_A}}{Z_A}.
\end{equation}
\new{Now,} how does Eq.~\eqref{eq:nonin} change when we introduce local (and potentially strong) interactions? Can we identify the state of a subsystem with some thermal state? How different is it from $\frac{e^{-\beta H_A}}{Z_A}$? This general question sometimes goes under the name of \emph{locality of temperature} \cite{kliesch2014}.

There are (to the author's knowledge) two different but related answers to this: the idea of \emph{local indistinguishability} and also the notion of \emph{Hamiltonian of mean force}. We now explain both of them, elaborate on their significance \new{for thermodynamics}, and also give a proof of the simplest instance of the first (in 1D). %These results largely simplify the study of local properties of thermal states, in that they show how local properties are largely independent of the bulk and can be calculated just by computing a small subsystem \cite{kliesch2014,alhambra2021locally}.

%%%%%
%%%%%

\subsection{Local indistinguishability}\label{sec:localind}

Given the above discussion on the decay of correlations, we expect that the state of a local subsystem will not depend much on the parts that are far away enough from it. A possible way to phrase this is that the local marginal $\rho_A$ is indistinguishable from the marginal of a much smaller thermal state, with a Hamiltonian that acts only in the vicinity of $A$. We now make this intuition precise. 

\new{Let us refer to partitions into $ABC$ such as those in Fig. \ref{fig:shield} or Fig. \ref{fig:shield1D}}, and write the Hamiltonian with the following terms:
\begin{equation}
 H= H_A +H_{AB} +H_B +H_{BC} +H_C.
 \end{equation}
We now have the full thermal state $\rho$, as well as a thermal state supported on $A,B$ defined as
\begin{equation}
   \rho_0^{AB} = \frac{e^{-\beta( H_A + H_B +H_{AB})}}{Z_{AB}},
\end{equation}
that is, without the terms in $H$ that have support in $C$. One can also think of this as the marginal of the thermal state $\rho_0^{AB}\otimes \rho_0^C \equiv e^{-\beta (H_{AB}+H_C)}/Z_{AB}Z_C$ in which we have removed the interactions $H_{BC}$ between $AB$ and $C$. Notice that $\rho_0^{AB} \neq \rho^{AB}$ due to the presence of $H_{BC}$. This is, however, just a small local term. 

The main idea is that if $B$ is large enough, these two states are almost indistinguishable on $A$. Let us assume that the connected correlations from Eq. \eqref{eq:corrdecay} decay with function $f(\text{dist}(C,D))$. Then, the following upper bound holds for some constant $K>0$ \cite{brandao2019finite}  
\begin{align}\label{eq:local1norm2}
    \vert \vert &\tr_{BC} [\rho]-\tr_{B}[ \rho_0^{AB}] \vert \vert_1 \\ \nonumber & \le K \vert \partial_C \vert \left( f(\text{dist}(A,C)) + e^{-\Omega( \text{dist}(A,C))} \right).
\end{align}
The first term in the RHS comes from the decay of correlations assumption. The second comes from using the QBP technique in Sec. \ref{sec:QBP}. The exponential decay of this quantity thus holds whenever both the correlations decay fast enough, and Lieb-Robinson bounds hold. An alternative proof for high temperatures using the cluster expansion can also be found in \cite{kliesch2014}.

A straightforward consequence of this is that we do not need to know the whole state to compute local quantities. If we care about some kind of local order parameter, or want to compute currents or else between some part and its surroundings, we can calculate them without having to diagonalize a huge matrix of size $\exp(N)$, but rather just focus on a much smaller region. \new{This is particularly useful in translation-invariant systems.} %This has various implications for the construction of classical and quantum algorithms, as we  describe in Sec. \ref{sec:algorithms}.

\subsubsection{Proof in 1D}\label{sec:local1D}

We now show the full proof of this result in the case of one dimension. The more general one, however, is essentially the same and can be found in \cite{brandao2019finite}. It uses previously mentioned results, and shares some steps and ideas that appear in other fundamental questions including the proof of the absence of phase transitions in 1D \cite{araki1969,Harrow_2020} or of decay of correlations \cite{araki1969,bluhm2021exponential}. It will also be a key ingredient in the algorithm of Sec. \ref{sec:algo1D}.

\begin{figure}[t]
\centering
\includegraphics[width=0.95\linewidth]{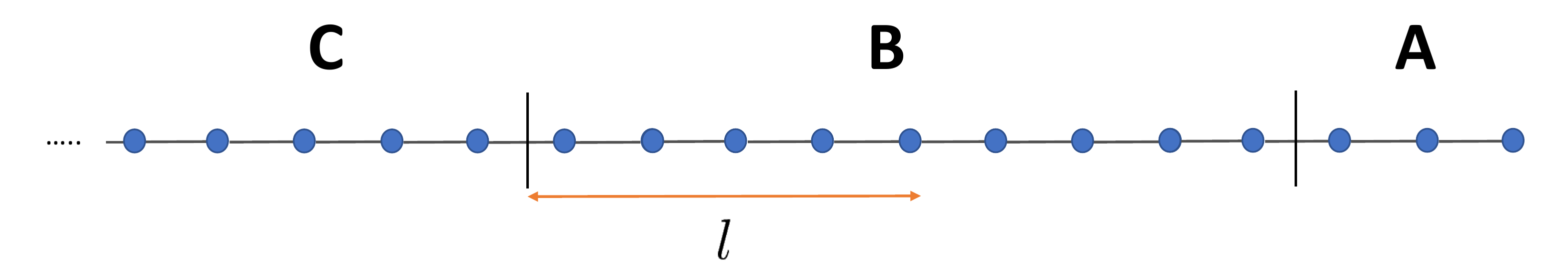}.
\caption{Choice of regions for the proof in Sec. \ref{sec:local1D}, and depiction of the distance $l$ on which the operator $E_{BC}^l$ acts within $B$.}\label{fig:chain}
\end{figure}

We focus on the restricted setting of a chain, that we divide into three parts $A,B,C$, such that $B$ is in the middle and $A$ is a small subsystem at the end of the chain, as in Fig. \ref{fig:chain}.
The aim is a small upper bound on
\begin{equation}\label{eq:local1norm}
    \vert \vert \tr_{BC} [\rho]-\tr_{B}[ \rho_0^{AB}] \vert \vert_1 = \max_{\vert \vert M_A \vert \vert \le 1} \vert \tr[M_A (\rho-\rho_0^{AB}\otimes \rho_0^C)] \vert,
\end{equation}
where $M_A$ has support on $A$ only, and the equality comes from the definition of the $1$-norm. Now, let us define the following two operators
\begin{itemize}
    \item 
    $E_{BC}=e^{\beta H}e^{-\beta (H-H_{BC})}$, 
\item $E^{l}_{BC}=e^{\beta (H^{l}_C+H^{l}_B+H_{BC})}e^{-\beta (H^{l}_B+H^{l}_C)}$, where $H^{l}_B$ and $H^{l}_C$ are the terms of $H_B$ and $H_C$ that are a distance at \new{most} $l$ from the boundary terms $H_{BC}$. 
\end{itemize}
%The second one $E^{l}_{BC}$ is the same as $E_{BC}$ but restricting the terms that appear in the exponents to be in the vicinity of $H_{BC}$. 
The parameter $l$ is free, so we can choose to our convenience. We refer now to the result from \cite{araki1969} in Eq. \eqref{eq:araki1} and \eqref{eq:araki2}, from which it follows that
\begin{align} \label{eq:araki3}
 \vert \vert E_{BC} \vert \vert &\le C_1 \\ \label{eq:araki4}
 \vert \vert  E_{BC} - E^{l}_{BC} \vert \vert &\le C_2 \frac{q^{1+l}}{(l+1)!}
\end{align}
That is, the operator $E_{BC}$ has bounded norm and, since we can approximate it by $E^{l}_{BC}$ with some $l< \text{dist}(A,C)$, its support on region $A$ is super-exponentially suppressed in $l$ (due to the factorial, which always dominates over $q^l$). In what follows, we choose $l=\vert B \vert/2$. Notice that by definition $\rho_0^{AB}\otimes \rho_0^C  = \frac{Z}{Z_{AB}Z_C} \rho E_{BC} $. 

Let us define $M_A^*$ to be the operator that optimizes the RHS of Eq. \eqref{eq:local1norm}. With the triangle inequality we can write
\begin{align}
    &\vert \tr[M^*_A (\rho-\rho_0^{AB}\otimes \rho_0^C)] \vert \\ \nonumber &\le   \left \vert \tr[M^*_A (\rho-\frac{Z}{Z_{AB}Z_C} \rho E^{l}_{BC} )] \right \vert \\ &\nonumber +  \left \vert \tr[M^*_A (\frac{Z}{Z_{AB}Z_C} \rho E^{l}_{BC} -\rho_0^{AB}\otimes \rho_0^C)] \right \vert.
\end{align}
Let us now upper-bound these two terms independently. The second can be bounded with Eq.~\eqref{eq:araki4} and H\"older's inequality applied twice.
\begin{align}
    &\left \vert \tr[M^*_A \frac{Z}{Z_{AB}Z_C} \rho E^{l}_{BC} -\rho_0^{AB}\otimes \rho_0^C)] \right \vert  \\ &= 
    \left \vert \tr[M^*_A \frac{Z}{Z_{AB}Z_C} \rho(E^{l}_{BC}-E_{BC}) ] \right \vert
   \\ & \le \frac{Z}{Z_{AB}Z_C} \vert \vert M^*_A \vert \vert \vert \vert \rho \vert \vert_1 \vert  \vert E^{l}_{BC}-E_{BC} \vert \vert
  \\& \le \frac{Z}{Z_{AB}Z_C} \times C_2 \frac{q^{1+l}}{(1+l)!}.
\end{align}
 Given Eq. \eqref{eq:mineq1}, $\max\{\frac{Z}{Z_{AB}Z_C},\frac{Z_{AB}Z_C}{Z}\}\le e^{\beta \vert \vert H_{BC} \vert \vert}$ , which is a constant that only depends on $\beta,k,J$. Thus, this second term is super-exponentially suppressed in $\vert B \vert$. 
 
 For the first term, we require the decay of correlations property Eq. \eqref{eq:corrdecay} \new{(which holds in 1D under the assumption of translation invariance)}. Since $l=\vert B \vert/2$,
 \begin{align}
    & \left \vert \tr[M^*_A \rho E^l_{BC} ]-\tr[M^*_A \rho] \tr[ \rho E^l_{BC} ] \right \vert \\ & \le K e^{-\frac{\vert B \vert}{2 \xi}} \vert \vert E^l_{BC} \vert \vert \le  2 K C_1 e^{-\frac{\vert B \vert}{2 \xi}},
 \end{align}
 where for the last inequality we used $\vert \vert E^l_{BC} \vert \vert \le \vert \vert E_{BC} \vert \vert+ \vert \vert E^l_{BC} -E_{BC}\vert \vert \le 2 C_1 $, which holds for sufficiently large $l$. We can now write
 \begin{align}
    & \left \vert \tr[M^*_A (\rho-\frac{Z}{Z_{AB}Z_C} \rho) E^{l}_{BC} ] \right \vert  \\ \nonumber &\le \left\vert \tr[M^*_A \rho] - \frac{Z}{Z_{AB}Z_C} \tr[M^*_A \rho] \tr[\rho E^{l}_{BC} ] \right \vert +  2 K C_1 e^{-\frac{\vert B \vert}{2 \xi}} \\ \nonumber &
     \le \left(1 - \frac{Z}{Z_{AB}Z_C} \tr[ \rho E^{l}_{BC} ] \right) +  2 K C_1 e^{-\frac{\vert B \vert}{2 \xi}},
 \end{align}
 where we used the triangle inequality in the first line, and H\"older's inequality $\tr[M^*_A \rho] \le \vert \vert M^*_A \vert \vert \le 1$ to get to the second. Finally, we can use Eq. \eqref{eq:araki4} again after another application of H\"older's inequality
 \begin{align}
     \vert \tr[\rho E^l_{BC}]-\tr[\rho E_{BC}] \vert \le \vert \vert  E_{BC} - E^{l}_{BC} \vert \vert \le C_2 \frac{q^{1+l}}{(1+ l)!},
 \end{align}
 and since $\tr[\rho E_{BC}]= \frac{Z}{Z_{AB}Z_C}\le e^{\beta \vert \vert H_{BC} \vert \vert}$ we obtain
 \begin{align}
     &\left \vert \tr[M^*_A (\rho-\frac{Z}{Z_{A B}Z_C} \rho E^{l}_{BC} )] \right \vert \\ &\le  C_2 e^{\beta \vert \vert H_{BC} \vert \vert} \frac{q^{1+l}}{(1+ l)!} +  2 K C_1 e^{-\frac{\vert B \vert}{2 \xi}}.
 \end{align}
 
 This finishes the proof. Putting everything together, we see that we have upper-bounded our target quantity in Eq. \eqref{eq:local1norm} by a small number related to the error term in the decay of correlations and Araki's result. Without writing the constants explicitly, and just on the leading exponential error, the final result is
 \begin{equation}\label{eq:localresult}
    \vert \vert \tr_{BC} [\rho]-\tr_{B}[ \rho_0^{AB}] \vert \vert_1 \le e^{-\Omega (\vert B \vert)},
\end{equation}
where $\Omega(x)$ is defined in Sec. \ref{sec:bigO}.% big-O notation for a function that grows at least as fast as $x$.

For simplicity, we have only dealt with the case of a 1D chain, where $A$ is at the end of it. To generalize the proof, one just needs to define an analogous partition $ABC$ in higher dimensions (see Fig. \ref{fig:shield} for 2D) and then remove all the different interaction terms from $H_{BC}$ one by one. Here, we have done it with the operator $E_{BC}$, but this can also be done with the (suitably defined) QBP operator $O_A$ from Sec.~\ref{sec:QBP}, and the result Eq. \eqref{eq:local1norm2} is essentially unchanged.

%%%%%%%%
%%%%%%%%
 
\subsection{Hamiltonian of mean force}\label{sec:meanforce}

The state $\rho_A$ is obviously the thermal state of \emph{some} Hamiltonian on $A$, since we can always define
\begin{equation} \label{eq:HMF}
    \tilde H _A \equiv \beta^{-1} \log \tr_{\setminus A}[e^{-\beta H}],
\end{equation}
which is \new{in general} different from $H_A$.
This is the so-called Hamiltonian of mean force \cite{miller2018hamiltonian}.  \new{Notice that it can be defined up to some additive constant chosen at will.} 

How does this Hamiltonian compare to the ``bare" Hamiltonian $H_A$, which disregards the interactions between $A$ and the rest? That is, we would like to understand the norm and locality of the operator $\Phi_A \equiv \tilde H_A -H_A$.
This turns out to be a difficult problem, \new{very much related to both the decay of mutual information and of the conditional mutual information from Sec. \ref{sec:correlations}.}

%\new{We now briefly describe a known result for high temperatures from \cite{Kuwahara_2020_Markov}, whose proof involves involves the connected cluster expansion applied to Eq. \eqref{eq:HMF}.}
\new{One potential result is as follows. 
Since the interactions are local, it makes sense that, if the size of $A$ is much larger than the number of nearest neighbours $k$, most of the weight of $\Phi_A$ is localized around its boundary with the rest of the system, of size $\vert \partial A \vert$.} %More precisely: can we approximate $\Phi_A$ with another operator $\Phi_A^l$ that only has support on sites a distance $l$ away from the boundary? }

\begin{figure}[t]
\centering
\includegraphics[width=0.5\linewidth]{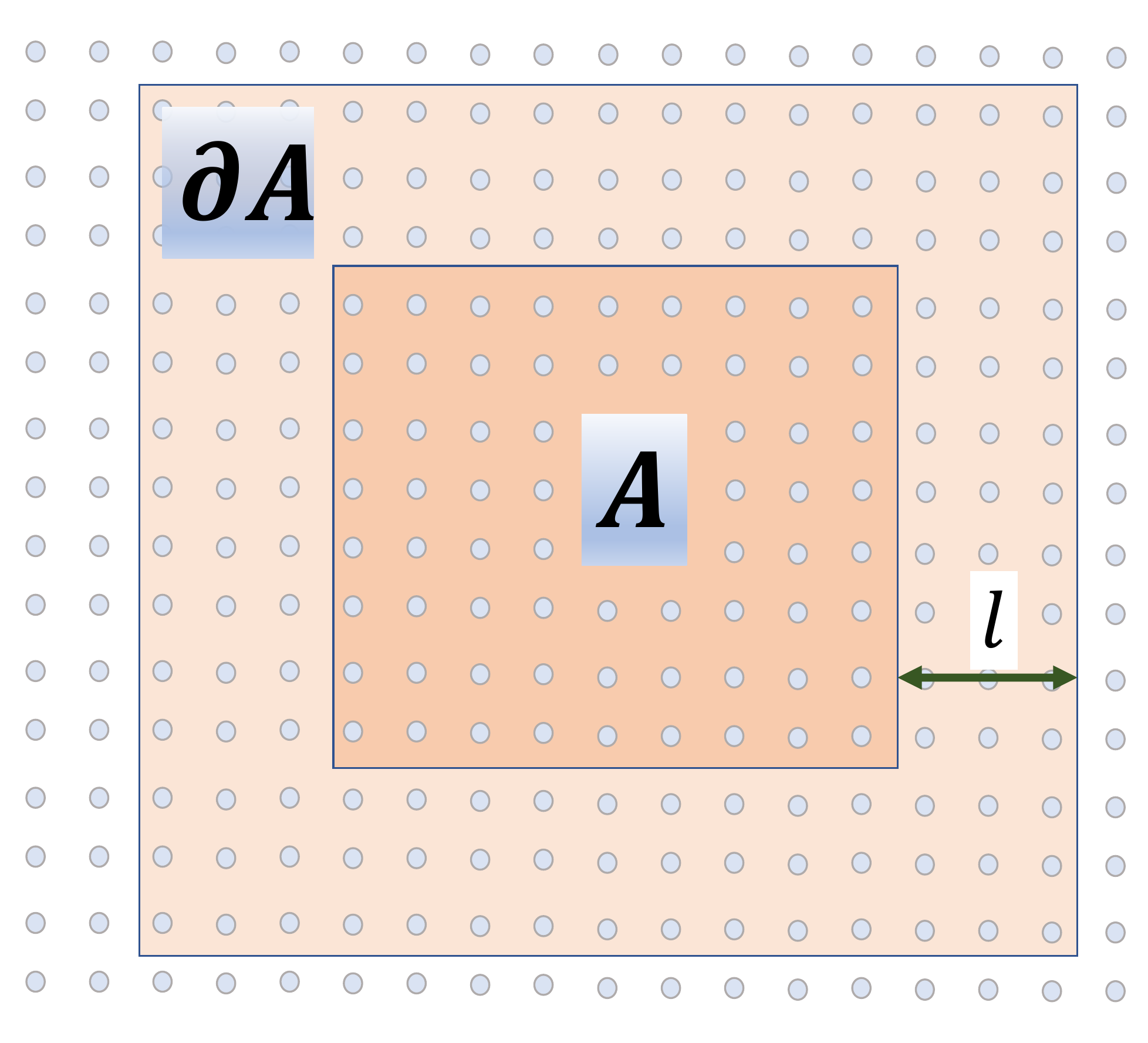}.
\caption{Illustration of the regions of the Hamiltonian of mean force. The correction term $\Phi_A$ is exponentially well approximated by $\Phi_A^l$, which has support on the region $\partial A$ only.}
\label{fig:meanforce}
\end{figure}

If the size of $A$ is much larger than the number of nearest neightbours $k$, most of the weight of $\Phi_A$ is localized around its boundary with the rest of the system, of size $\vert \partial A \vert$. The precise question is: can we approximate $\Phi_A$ with another operator $\Phi_A^l$ that only has support on sites a distance $l$ away from the boundary? Using a non-commutative cluster expansion, as in Sec. \ref{sec:CMI}, Theorem 2 in \cite{Kuwahara_2020_Markov} shows that, for any temperature $\beta$ above a threshold one $\beta^*_{\text{NC}}>0$, one can define a $\Phi_A^l$ such that
\begin{equation}
    \vert \vert \Phi_A -\Phi_A^l \vert \vert \le \frac{e}{2\beta}(\beta/\beta^*_{\text{NC}})^{l/k} \vert \partial A \vert.
\end{equation}
That is, $\Phi_A$ can be exponentially well approximated with an operator localized around the boundary. See Fig. \ref{fig:meanforce} for an illustration.

A similar result is expected to hold in 1D, but this is a so far open problem. See \cite{trushechkin2021open} for a recent overview on this topic for a different set of models, and its implications.

%\new{A way to do this is suggested by \cite{Kuwahara_2020_Markov} where it is shown that if the cluster expansion of $\log \tr_{\setminus A}[e^{-\beta H}]$ converges,  one could define a $\Phi_A^l$ with support at most a distance $l$ from the boundary that approximates $\Phi_A$ exponentially well in $l$.}
%\begin{equation}
%    \vert \vert \Phi_A -\Phi_A^l \vert \vert \le \mathcal{O}\left( \vert \partial_A \vert \left(\frac{\beta}{\beta^*}\right)^m \right).
%\end{equation}
%That is, $\Phi_A$ can be exponentially well approximated with an operator localized around the boundary.
%\new{See Fig. \ref{fig:meanforce} for an illustration. However, this is currently an open question since the convergence of that cluster expansion for operators has currently not been proven, and does not follow straightforwardly from the discussion of Sec. \ref{sec:connectedclusters}, similar to Sec. \ref{sec:CMI}.}

%See also \cite{trushechkin2021open} for a recent overview on this topic for a different set of models, and its implications for thermodynamics, \new{which we now briefly discuss}.

%%%%%%%%%%%%%%%%%
%%%%%%%%%%%%%%%%%

\subsection{Non-equilibrium thermodynamics with strong coupling}\label{sec:strongcoupling}

\new{The ideas of this section may help understand how a priori complex non-equilibrium thermodynamic processes may be tractable in practice. In many situations of interest, the starting point at $t=0$ is an equilibrium state with time-dependent Hamiltonian
\begin{equation}
    H(t)=H_S(t)+ H_B + H_I.
\end{equation}
This includes a system Hamiltonian $H_S(t)$, a bath Hamiltonian $H_B$, and an interaction $H_I$ between them \footnote{This interaction might also depend on time, which here we do not consider for simplicity.}. The driving of $H_S(t)$ for $t>0$ takes the system away from equilibrium, and different thermodynamic quantities can then be studied.}

\new{Textbook thermodynamics are typically centered around macroscopic systems such as gases, where one can take the weak coupling limit $H_I \ll H_S, H_B$. In many body quantum systems, such as the ones considered here, this limit may no longer apply, which creates a number of difficulties for thermodynamic considerations (see e.g. \cite{talkner2020colloquium,miller2018hamiltonian,Strasberg2019}). These, however, can be dealt with if one considers the effective state on the system 
\begin{equation}
    \rho^S(t) = \frac{\tr_B[e^{-\beta H(t)}]}{\tr_{SB}[e^{-\beta H(t)}]} \equiv \frac{e^{-\beta \tilde H_S(t)}}{\tilde Z_S(t)},
\end{equation}
where it is convenient to define the Hamiltonian of mean force $\tilde H_S(t)$ such that  $\tilde Z_S(t) \equiv \frac{\tr_{SB}[e^{-\beta H(t)}]}{\tr_B[e^{-\beta H_B}]}$. This way, for instance, the non-equilibrium free energy (and thus the second law) is defined as
\begin{equation}
    F_S(t) = \tr_S[\tilde H_S(t) \rho^S(t)]- \frac{1}{\beta}S(\rho^S(t)).
\end{equation}
See e.g. \cite{Strasberg2019} for details. These quantities may be difficult to calculate. While $\tilde H_S(t)$ may be inferred from the system alone, $\tilde Z_S(t)$ may depend on the system but also potentially on its relation to the whole bath. }

\new{Consider the rather general situation in which the system to be a small part of a lattice Hamiltonian of a Gibbs state with short-ranged correlations, and the bath to be the rest of the lattice. In that case, the situation simplifies dramatically. First, the discussion from Sec. \ref{sec:meanforce} suggests that typically the Hamiltonian of mean force $\tilde H_S$ does not necessarily depend on the whole bath, but has some corrections which only depend on the region around the boundary between $S$ and $B$. }

\new{Then, for the effective partition function $\tilde Z_S$, the same fact follows from local indistinguishability. To show this, notice that
\begin{align}
 \frac{\tilde Z_S(t)}{Z_S(t)} \equiv & \frac{\tr_{SB}[e^{-\beta H(t)}]}{\tr_{SB}[e^{-\beta H_S(t)} \otimes e^{-\beta H_B}]} \\ =& \nonumber \tr_{SB}[O^\dagger_{I}O_{I} \frac{ e^{-\beta H_S(t)} \otimes e^{-\beta H_B}}{\tr_{SB}[e^{-\beta H_S(t)}   \otimes e^{-\beta H_B}]}]
 \\ =&  \nonumber  \tr_{SB}[(O^l_{I})^\dagger O^l_{I} \frac{e^{-\beta H_S(t)}  \otimes e^{-\beta H_B}}{\tr_{SB}[e^{-\beta H_S(t)} \otimes e^{-\beta H_B}]}] \nonumber \\ &+ e^{\mathcal{O}(\beta \norm{H_I(t)}-l)} 
\end{align}
where $O_{I}$ is the belief propagation operator from Eq. \eqref{eq:QBPoperator} with $A=H_I$ and $O^l_{I}$ is its local approximation in Eq. \eqref{eq:QBPoperator2}. Defining $B_l$ to be the region of the bath that is at most a distance $2l$ away from $S$, we have that 
\begin{align}
 \frac{\tilde Z_S(t)}{Z_S(t)} =&   \tr_{SB_l}[(O^l_{S})^\dagger O^l_{S} \frac{e^{-\beta H_S(t)}  \otimes e^{-\beta H_{B_l}}}{\tr_{SB_l}[e^{-\beta H_S(t)} \otimes e^{-\beta H_{B_l}}]}] \nonumber \\ & + e^{\mathcal{O}(\beta \norm{H_I}-l)}.
\end{align}}

\new{This shows that the effective partition function can be approximated to multiplicative error $\epsilon$ by computing the expectation value of $O^{\dagger}_I O_I$ on the system and a region of the bath a distance $l= \log(\epsilon^{-1}) + \mathcal{O}( \beta \norm{H_I}) $ away from the small system. In $D$ dimensions, assuming $\norm{H_I} = \mathcal{O}(1)$, the computational cost of exact diagonalization is $e^{\mathcal{O}(l)}=e^{\mathcal{O}(\log^{D}\epsilon^{-1})}$, independent of the size of the bath.}

%%%%%%%%%%%%%%%%%
%%%%%%%%%%%%%%%%%

\section{Statistical properties}\label{sec:statistical}

We now explain and prove some important statistical features of thermal states. These are central statements of the field of statistical physics and characterize the ensembles involved: the thermal (or canonical) and the microcanonical, as well as the grand canonical or others, when relevant. In contrast to the results of other sections, all those shown here (as well as their proofs) apply equally to classical models. 

\subsection{Measurement statistics and concentration bounds}\label{sec:concentration}

In Sec. \ref{sec:correlations} we saw how in many instances of thermal states, \new{ in particular for those in the ``one phase region"}, the different subsystems \new{tend not to} have strong correlations. This has a number of consequences, and we now explore an important one that shows that their large-scale statistical properties resemble those of non-interacting/statistically independent systems. These are \emph{concentration bounds}, akin to the (perhaps more widely known) central limit theorem. 

The setting is as follows: let us consider a $k$-local observable $A=\sum_j A_j$, such that $A_j$ has support on at most $k$ sites. The best example is the energy, but also other properties like magnetization $\sum_j \sigma_j^Z$. 

The expectation value of any such observable can be thought of as a macroscopic property of the system (such as the average magnetization of the material). While we expect that there will be thermal fluctuations around that average value, our intuition from thermodynamics tells us that any such large-scale property should have a definite value, almost free of fluctuations. This is due to one of the most basic ideas from probability theory: the measurement statistics of sums of independent random variables greatly concentrate around the average. The main conclusion is that if we measure an observable $A$ on a thermal state, the outcome will be very close to the average $\langle A \rangle_\beta$ with overwhelmingly large probability. That is, the distribution 
\begin{equation}
    P_{A,\beta}(x)= \tr[\rho \delta(x-A) ].
\end{equation}
 which is the probability of obtaining outcome $x$ when measuring $A$, is highly peaked around the average $\langle A \rangle_\beta = \tr[\rho_\beta A]$. 
This has important implications for the validity of thermodynamic descriptions of these systems, in that averaged macroscopic quantities characterize the large system of many particles whose properties we do not know with any certainty. 

\new{In the theory of probability, there are various types of results characterizing distributions comprised of many independent (or close to independent) variables}. Their proofs most often involve constraining  the characteristic function $\langle e^{\lambda A} \rangle_\beta$, where $\lambda$ may be real or imaginary. We now describe some of them.

\subsubsection{Chernoff-Hoeffding bound}

This is a concentration bound that reads
\begin{align}\label{eq:hoeffding}
   P_{A,\beta}(\vert x - \langle A \rangle_\beta \vert > \delta) \le 2 \exp \left ( - \frac{\delta^2}{4 c \bar A} \right),
\end{align}
where $\bar A \equiv \sum_j \vert \vert A_j \vert \vert$. \new{Thus} if $
\delta^2 
\gg c \bar A$, the probability of measuring $A$ to be away from $\langle A \rangle_\beta$ by at least $\delta$ is \new{exponentially} small. The most common proof technique is via a bound on the characteristic function of the form
\begin{equation}\label{eq:hoeffding2}
    \log \langle e^{\tau (A-\langle A \rangle_\beta)} \rangle_\beta \le c \tau^2 \bar A,
\end{equation}
for some \new{$\mathcal{O}(1)$ constant $c>0$ (which might depend on $\beta$ and the parameters of the Hamiltonian) and a wide enough range of $\tau$}. From this it follows that 
\begin{align}
  &   P_{A,\beta}( x - \langle A \rangle_\beta > \delta) = \int_{x - \langle A \rangle_\beta > \delta} \tr[\rho \delta(x-A) ]\\ & = \int_{x - \langle A \rangle_\beta > \delta} \tr[\rho e^{\tau (A-\langle A \rangle_\beta)} e^{-\tau (A-\langle A \rangle_\beta)} \delta(x-A)  ] \\ 
     &\le e^{-\tau \delta} \tr[\rho_\beta e^{\tau (A-\langle A \rangle_\beta)}] \le \exp \left(-\tau \delta + c \tau^2 \bar A \right).
\end{align}
One can follow the same steps for the range $ \langle A \rangle_\beta -x > \delta$. Then, choosing $\tau=\delta/ (2 c \bar A)$ yields Eq. \eqref{eq:hoeffding}.

This result can be very easily shown for independent random variables or for independent spins. For interacting spins, Eq. \eqref{eq:hoeffding2} was shown in \cite{Kuwahara_2020_Gaussian} with the cluster expansion technique from Sec. \ref{sec:connectedclusters}, which holds for all dimensions and all temperatures $\beta < \beta^*$. \new{To see this, notice that a bound of the form of Eq. \eqref{eq:hoeffding2} follows from proving the convergence of the expansion of $\log \langle e^{\tau (A-\langle A \rangle_\beta)} \rangle_\beta$ to second order.}
The main result from \cite{Anshu_2016} proves a slightly weaker version of Eq. \eqref{eq:hoeffding} with a different technique, only assuming the decay of correlations from Sec. \ref{sec:corrdecay}.

\subsubsection{Large deviation bound}

A related important type of concentration bound is given by large deviation theory. This is the branch of probability theory concerned with understanding the likelihood of very rare events, and has a long history as one of the most important mathematical frameworks for studying statistical physics. For instance it gives a way of describing the equilibrium properties of large ensembles (as is also the case here), or for predicting the long-time behaviour of non-equilibrium processes such as Brownian motion. See \cite{Touchette2009} for an excellent overview of the main results and their consequences for classical systems.

The basic idea is that given any set of measurement outcomes $\mathcal{A}$, we would like to identify whether there always exists a \emph{rate function} $I_{\mathcal{A}}$ such that
\begin{equation}
    \lim_{N \rightarrow \infty} -\frac{\log P_{A,\beta}(x \in \mathcal{A})}{N}= I_\mathcal{A}. 
\end{equation}
If this is the case, the dominant behaviour of $P_{A,\beta}(x \in \mathcal{A})$ is essentially a decaying exponential $P_{A,\beta}(x \in \mathcal{A}) \simeq e^{- N I_\mathcal{A} + o(N)}$, unless $I_\mathcal{A}=0$. This means that, in the thermodynamic limit, the measurement statistics of $A$ are extremely peaked around the points where the rate function vanishes $I_\mathcal{A}=0$. 

This is slightly stronger than the Chernoff-Hoeffding inequality, in that it can in principle give an exact expression of the probability distribution for large enough $N$. However, we do not always know how large an $N$ is ``enough'', and for finite $N$, it often does not give an expression as explicit as Eq. \eqref{eq:hoeffding}. 

Again, the proof strategy most often involves the characteristic function. In particular, the G\"artner-Ellis theorem states that a sufficient condition is that the function
\begin{equation}
    g(\tau) =\lim_{N \rightarrow \infty} \frac{\log \langle e^{\tau A} \rangle_\beta}{N}
\end{equation}
exists and is differentiable. This has been shown using the cluster expansion in \cite{LargeDevCluster2004} for $1$-local observables, and upper bounds on the rate for general observables have been shown using the locality estimates from Sec. \ref{sec:localityestimates} in \cite{Lenci2005,HiaiLarge2007}. The full large deviation principle was shown in 1D in \cite{Ogata2010}. An alternative proof can be found in \cite{OgataLucBellet2011}.

\subsubsection{Berry-Esseen theorem}

\new{
Another interesting probability theory result is the Berry-Esseen theorem \cite{berry1941accuracy,esseen1942liapounoff,brandao2015equivalence}, which can be thought of as a refinement of the central limit theorem for a finite sample size (which in this case is the system size $N$). 
Let us define the cumulative distribution function
\begin{align}
    F(x)= \int_{-\infty}^x  P_{A,\beta}(x) \text{d}x
\end{align}
as well as the equivalent for a Gaussian with the same average and variance
\begin{align}
    G(x) =\frac{1}{\sigma_A \sqrt{2\pi }} \int_{-\infty}^x e^{\frac{-(y-\langle A \rangle_\beta}{ 2\sigma^2_A}},
\end{align}
where $\sigma^2_A= \langle A^2 \rangle_\beta-\langle A \rangle_\beta^2$. The distance between these two functions is bounded by Esseen's inequality \cite{feller1991introduction}, which states that, for all $T >0$
\begin{align}
    \Delta &\equiv \max_x \vert F(x) - G(x) \vert \\ &\le \frac{18}{\sqrt{2 \pi^3} T}+ \frac{1}{\pi} \int_0^T \frac{\left \vert e^{-\frac{t^2}{2}}-\langle e^{i t \frac{A -\langle A \rangle_\beta}{\sigma_A}} \rangle \right \vert}{t} \text{d}t. \label{eq:BEerror}
\end{align}
That is, the right hand side is small if the characteristic function inside the integral is close to a Gaussian for rather long times $t$.}

\new{This can be shown by bounding the logarithm of the characteristic function $\langle e^{i t \frac{A -\langle A \rangle_\beta}{\sigma_A}} \rangle= \tr[e^{i t \frac{A -\langle A \rangle_\beta}{\sigma_A}} \frac{e^{-\beta H}}{Z}]$ with the cluster expansion from Sec. \ref{sec:connectedclusters}. Assuming $h=\mathcal{O}(1)$, for short times $t/\sigma_A \le t^*$, with $t^*$ some $\mathcal{O}(1)$ constant, it can be shown that it is close to the second order Taylor expansion as
\begin{align}
    \left \vert \log \langle e^{i t \frac{A -\langle A \rangle_\beta}{\sigma_A}} \rangle - \frac{-t^2}{2} \right \vert \le \mathcal{O} \left ( \frac{N t^3}{\sigma^3} \right),
\end{align}
so that 
\begin{align}
 \langle e^{i t \frac{A -\langle A \rangle_\beta}{\sigma_A}} \rangle = e^{-\frac{t^2}{2}+\mathcal{O} \left ( \frac{N t^3}{\sigma^3} \right)}.
\end{align}
To prove this, see for instance Theorem 13 in \cite{Wild_2023}.
This allows us to bound the integral in Eq. \eqref{eq:BEerror} choosing $T=\frac{t^* \sigma_A}{2}$, to achieve 
\begin{equation}
    \Delta \le \mathcal{O} \left ( \frac{1}{\sigma_A}+\frac{N}{\sigma_A^3} \right),
\end{equation}
which, considering that $\sigma_A= \Omega(\sqrt{N})$, means that $\Delta = \mathcal{O}(N^{-1/2})$. This means that the cumulative functions $F(x)$ and $G(x)$ become increasingly similar with system size, which shows that the probability $P_{A,\beta}(x)$ approaches a Gaussian in the thermodynamic limit. }

\new{A different proof starting from the assumption of decay of correlations, can be found in \cite{Brandao_2015G}.}

%%%%%%%%%%
%%%%%%%%%%

\subsection{Equivalence of ensembles}

We now prove an important statement in the study of statistical physics, which goes back all the way to Boltzmann and Gibbs. In large systems, the average macroscopic properties of both the thermal or canonical state, and of the microcanonical ensemble, are essentially the same. This means that both canonical and ergodic averages coincide in the thermodynamic limit, and shows that the particular ensemble used for calculations does not necessarily matter.

There are various similar statements in the literature \cite{Lima1971,Lima1972,MullerAdlamd2015,brandao2015equivalence,Tasaki2018,Kuwahara_2020_ETH,Kuwahara_2020_Gaussian}, but the proof that we now show follows that of \cite{Tasaki2018,Kuwahara_2020_ETH,Kuwahara_2020_Gaussian} and relies on the Chernoff-H\"offding bound. Let us define the extensive observable $A=\sum_j A_j $ (such as e.g. the total magnetization $\sum_{j}^N \sigma_j^Z $) with thermal/canonical average $\langle A \rangle_\beta$ which for simplicity we will set to $\langle A \rangle_\beta=0$, while the microcanonical average is 
\begin{equation}
    \langle A \rangle_{E,\Delta}= \frac{1}{D_N(E,\Delta)}\sum_{E_j \in (E-\Delta, E)} \langle E _j \vert A \vert E_j \rangle,
\end{equation}
where $E$ is the energy and $\Delta$ the width of the microcanonical window (which might depend on $N$), and $\vert E_j \rangle$ is the energy eigenstate of energy $E_j$. $D_N(E,\Delta)$ is a normalization constant counting the number of eigenstates within the window. This motivates the following probability distribution
\begin{align}
    P_{E,\Delta}(x)=\frac{1}{D_N(E,\Delta)}\sum_{E_j \in (E-\Delta, E)}\delta(x-\langle E_j \vert A \vert E_j \rangle ),
\end{align}
which gives the probability of measuring $x=\langle E_j \vert A \vert E_j \rangle$ when sampling eigenstates from the microcanonical ensemble.

First, we need to determine what is the energy that corresponds to temperature $\beta$ and thus characterizes the microcanonical ensemble. Given  the temperature $\beta$, the microcanonical energy $E_0$ is such that 
\begin{equation}\label{eq:microc}
    E_0(\Delta,\beta)\equiv
    \text{argmax}_{E} D_N(E,\Delta)e^{-\beta E}.
\end{equation}
Assuming that the width is significantly different than the energy scales of the system, $\Delta \ll \langle H \rangle_\beta$ (which is most typically the case), this roughly implies that $E_0$ is the energy of the microstates $\{ \vert E_j \rangle\}$ that have the dominant weight in the canonical ensemble (when the density of states is weighted by the factor $e^{-\beta E}$). We have written the dependence on $\beta,\Delta$ explicitly in Eq. \eqref{eq:microc}, but let us now drop them for simplicity of notation.

We start by upper bounding the $m$-th (even) moment of  $P_{E_0,\Delta}(x)$
\begin{align}
  &  \int_{-\infty}^{\infty} x^m P_{E_0,\Delta}(x) \\ \nonumber  &= \frac{1}{D_N(E_0,\Delta)} \sum_{E_j \in (E_0-\Delta, E_0)} \vert\langle E _j \vert A \vert E_j \rangle \vert^m \\ \le &  \frac{1}{D_N(E_0,\Delta)} \sum_{E_j \in (E_0-\Delta, E_0)} \vert\langle E _j \vert A^m \vert E_j \rangle \vert = \langle A^m \rangle_{E_0,\Delta}, \nonumber
\end{align}
where we we used the convexity of $x^m$ with $m$ even. The bound can easily be expressed in terms of a canonical average as, since $A^m$ is positive, 
\begin{align}
   & \langle A^m \rangle_{E_0,\Delta} =\frac{1}{D_N(E_0,\Delta)} \sum_{E_j \in (E_0-\Delta, E_0)} \langle E _j \vert A^m \vert E_j \rangle \\ & \le \frac{e^{\beta E_0}}{D_N(E_0,\Delta)} \sum_{E_j \in (E_0-\Delta, E_0)} e^{-\beta E_j} \langle E _j \vert A^m \vert E_j \rangle \\&\le \frac{e^{\beta E_0}}{D_N(E_0,\Delta)} \sum_{E_j \in (-\infty, \infty)} e^{-\beta E_j} \langle E _j \vert A^m \vert E_j \rangle \\& = \frac{Z e^{\beta E_0}}{D_N(E_0,\Delta)} \langle A^m \rangle_\beta.
\end{align}

The factor $\frac{Z e^{\beta E_0}}{D_N(E_0,\Delta)}$ can now be upper bounded using the definition of the microcanonical ensemble and the concentration bound. Let us define the following modified partition function $
    \tilde Z \equiv \sum_{\vert E_j-E_0 \vert \le \delta} e^{-\beta E_j}$. If we also set $\delta = K N^{1/2}$ with $K =\mathcal{O}\left(1\right)$ it follows from Eq. \eqref{eq:hoeffding} that 
\begin{align}
    \frac{\tilde Z}{Z}&=1-  P_{H,\beta}(\vert x- \langle H \rangle_\beta \vert \ge \delta) \\ & \ge 1- 2 \exp \left ( - \frac{\delta^2}{4 c J N} \right)\ge 1/2.
\end{align}
Now divide the energy range in the sum in equal parts of width $\Delta^* \equiv \min\{ \Delta, \beta^{-1}\}$, such that the largest energy of each interval is $E_{\nu}$, so that $E_{\nu+1}=E_{\nu}+\Delta^*$ and
\begin{align}
 \tilde Z   &\le \sum_{\substack{\nu \in \mathbb{Z}\\ \vert E_\nu - E_0 \vert  \le \Delta^* + \delta}} D_N(E_\nu,\Delta^*) e^{-\beta (E_\nu-\Delta^*)} \\ & \le  e^{\beta \Delta^*}\left( \frac{2\delta}{\Delta^*}+2 \right) \max_\nu D_N(E_\nu,\Delta^*) e^{-\beta E_\nu} \\ &\le \frac{1}{2}K' \frac{N^{1/2}}{\Delta^*} D_N(E_0,\Delta) e^{-\beta E_0},
\end{align}
with $K' =\mathcal{O}(1)$, where the last inequality follows from the fact that $D_N(E_0,\Delta)$ is monotonic on $\Delta$. We thus have $  \int_{-\infty}^{\infty} x^m P_{E_0,\Delta}(x) \le K' \frac{N^{1/2}}{\Delta^*}  \langle A^m \rangle_{\beta}$. To finish this part of the proof we bound  $\langle A^m \rangle_{\beta}$. It was shown in \cite{Kuwahara_2020_ETH,Kuwahara_2020_Gaussian} that the concentration inequality Eq. \eqref{eq:hoeffding} implies that
\begin{equation}\label{eq:moments}
     \langle A^m \rangle_{\beta} \le \left( 4 c \bar A \right)^{m/2}\left(\frac{m}{2} \right) !.
\end{equation}
For completeness, we reproduce the proof in Appendix \ref{app:usefullema}. We are now in a position to bound the tail of $P_{E,\Delta}(x)$ as
\begin{align}
   \nonumber & P_{E,\Delta}(x \ge x_0 )
   = \int_{x_0}^{\infty}   P_{E,\Delta}(x ) \text{d}x \le \frac{1}{x_0^m}  \int_{-\infty}^{\infty} x^m P_{E_0,\Delta}(x) \\& \le   K' \frac{N^{1/2}}{\Delta^*} \left(\frac{ 4 c \bar A }{x_0^2}\right)^{m/2}\left(\frac{m}{2} \right) ! 
    \le K' \frac{N^{1/2}}{\Delta^*} \left(\frac{ 4 m c \bar A }{x_0^2}\right)^{m/2}.
\end{align}
Thus, choosing $m=\lfloor{\frac{x_0^2}{4 c e \bar A}}\rfloor$. and repeating for $x \le -x_0$, leads to (let us now bring back the average $\langle A \rangle_\beta$ explicitly, previously taken to be zero)
\begin{equation}
     P_{E,\Delta}(\vert x - \langle A \rangle_\beta \vert \ge  x_0 ) \le  2 e K' \frac{N^{1/2}}{\Delta^*} \exp \left( - \frac{x_0^2}{8 c e \bar A} \right).
\end{equation}
We are almost done. We now bound the difference between canonical and microcanonical as
\begin{align}
& \vert   \langle A \rangle_{E,\Delta}- \langle A \rangle_\beta \vert \le \sum_{j} \frac{\vert \langle E_j \vert A \vert E_j \rangle - \langle A \rangle_\beta \vert}{D_{E_0, \Delta}} \\ &\le \int_{\vert x \vert \le \bar A} P_{E,\Delta}(x-\langle A \rangle_\beta)(x-\langle A\rangle_\beta)\text{d}x\\ & \le x_0 +2\bar A P_{E,\Delta}(\vert x - \langle A \rangle_\beta \vert \ge  x_0 ),
\end{align}
and so choosing $x_0 = \sqrt{8 c e \bar A} \log (4 \bar A e K' \frac{N^{1/2}}{\Delta^*})$, the fact that $\bar A \propto N$ yields, for some constant $K''$,
\begin{equation}
 \frac{1}{N}   \vert   \langle A \rangle_{E,\Delta}- \langle A \rangle_\beta \vert \le \frac{K'' \log \frac{N^{3/2}}{\Delta^*}} {N^{1/2}}
,\end{equation}
so that the difference vanishes in the thermodynamic limit. Notice that $\Delta^* \equiv \min \{\Delta, \beta^{-1} \}$, so that in principle even rather low temperatures and very small (up to exponentially small) microcanonical windows are allowed. This is the final result. It states that average properties are essentially the same, \new{provided that the average energy $E_0$ is determined by Eq. \eqref{eq:microc}}, and that the width $\Delta$ is not too small. The fact that it can be up to exponentially small in system size is rather strong, and related to weak statements of the eigenstate thermalization hypothesis (see  \cite{mori2016weak,Kuwahara_2020_ETH}).

%%%%%%%%%%%%%%%%
%%%%%%%%%%%%%%

\section{Algorithms and complexity of thermal states}\label{sec:algorithms}

When addressing specific problems in many-body physics, we would most often like to understand whether they are fundamentally complex or not, in the precise sense established by theoretical computer science. This can typically done in two complementary ways:
\begin{itemize}
     \item By showing that there exists an algorithm with a provable performance and run-time. Additionally, it is interesting if the algorithm can be explicitly constructed, and implemented in practice.
    \item By establishing that a problem, or a set of them, belong to or are complete or hard for a certain complexity class. 
\end{itemize}
This applies to both classical and quantum computation, and their respective complexity classes. 

Problems related to quantum thermal states can also be studied under this light. The relevant ones include most notably the estimation of the partition function, or the generation of either approximations to the thermal states (in quantum computers) or their classical representations (in classical computers). 

As an illustrative example of what can be proven, we start with a simple explicit algorithm that approximates the quantum partition functions in 1D in polynomial time \cite{kuwahara2018polynomialtime}. We then briefly review some other important known results about the hardness of approximating partition functions. The rest of the section includes an explanation of the current best tensor network results, which are provably efficient in a wide range of situations, and a short review of quantum algorithms for preparing thermal states.

\subsection{An efficient classical algorithm for the 1D partition function}\label{sec:algo1D}

Using some of the results from the previous sections, we now show that, assuming that $h,\beta =\mathcal{O}(1)$, and \new{that exponential decay of correlations holds}, we can efficiently approximate the partition function in 1D. This is done with an algorithm with runtime $ \text{poly}(N,\varepsilon^{-1})$ that outputs $Z'$, where
\begin{equation}\label{eq:approxZ1D}
\vert \log Z' - \log Z \vert \le \mathcal{O}\left ( \varepsilon\right).
\end{equation}
This section follows the result and proof strategy from \cite{kuwahara2018polynomialtime}, with some minor modifications.

In one dimension, let us consider the partial Hamiltonian $H_j=\sum_{i=1}^{j-1} h_i$, which includes the first $j-1$ interaction terms as counted from the left, starting from the left-most $h_1$. Then, define the partial partition function
\begin{align}\label{eq:partialZ}
    Z_i &=\tr[e^{-\beta(H_i+h_i)}] \\ &=\tr[O_{h_i}e^{-\beta(H_i)}O_{h_i}^\dagger] \equiv \tr[e^{-\beta(H_i)}A_{i}] ,
\end{align}
where $O_{h_i}$ is the quantum belief propagation from Sec. \ref{sec:QBP} and $A_{i}=O_{h_i}^\dagger O_{h_i}$. Now, rewriting Eq. \eqref{eq:partialZ} notice the simple iterative relation
\begin{equation}
  Z_i = Z_{i-1} \tr[\rho_i A_i],
\end{equation}
where $\rho_i= e^{-\beta H_i}/Z_{i-1}$. Thus we can write
\begin{equation}
    Z= d^{N} \prod_{i=1}^{\vert E \vert}  \tr[\rho_i A_i],
\end{equation}
where $Z \equiv Z_{\vert E \vert}$ and $d^N=Z_0$. The key now is to use results from Sec. \ref{sec:QBP} to approximate $A_i$, and local indistinguishability from Sec. \ref{sec:local1D}. Let $A_i^l \equiv (O^l_{h_i})^\dagger O^l_{h_i}$, so that 
\begin{align}
    \vert \vert A_i - A_i^l\vert \vert &=  \vert \vert A_i -O_{h_i}^\dagger O_{h_i}^l +O_{h_i}^\dagger O_{h_i}^l - A_i^l\vert \vert  \\ &\le 2 \vert \vert O_{h_i} \vert \vert \vert O_{h_i} - O_{h_i}^l \vert \vert
    \\ & \le  e^{ \mathcal{O}(\beta h)} e^{-\Omega(l)},\label{eq:boundai}
\end{align}
where in the first line we used the triangle inequality and in the second we used both Eq. \eqref{eq:QBPnorm} and \eqref{eq:QBPlocal}. Now, \new{let us label by $\Lambda_{l^*}$ to be the rightmost part of the chain of length $l^*$, with vertex set $V_{l^*}$ and in which $H_{i+1}$ has support. Choose $l^* \in \mathbb{R}$ so that $A_i^{l^*}$ has support in the right side of $V_{2l^*}$} and define $\rho_i^{(l_*)}= e^{-\beta H_{\Lambda_{l_*}}}/ \tr{e^{-\beta H_{\Lambda_{l_*}}}}$, where $H_{\Lambda_{l_*}}= \sum_{\supp{h_i} \in V_{l^*}} h_i$.

The expectation value can be approximated as
\begin{align}\label{eq:Abound}
&    \left \vert   \tr[\rho_i A_i]- \tr[\rho_i^{(2l_*)} A^{l_*}_i]  \right \vert \\ \nonumber &\le   \left \vert   \tr[\rho_i A_i]- \tr[\rho_i A^{l_*}_i]\right \vert +\left \vert \tr[\rho_i A^{l_*}_i]- \tr[\rho_i^{(2l_*)} A^{l_*}_i]  \right \vert
    \\ & \le  \vert \vert A_i - A_i^{l^*}\vert \vert + \vert \vert A_i^{l^*} \vert \vert \times \vert \vert \text{Tr}_{\setminus V_{l^*} }[\rho_i]-\text{Tr}_{V_{2l^*} \setminus V_{l^*} }[\rho_i^{(2l_*)}] \vert \vert_1.\nonumber
\end{align}
This follows from the triangle inequality. The partial trace $\setminus V_{l^*}$ is over the support of $\rho_i$ excluding \new{vertices $V_{l^*}$}. Eq. \eqref{eq:Abound} now has a form that we can upper bound. Since $\vert \vert A_i^{l^*} \vert \vert \le \vert \vert A_i- A_i^{l^*} \vert \vert + \vert \vert  A_i \vert \vert \le  e^{ \mathcal{O}(\beta h)} $, we can use Eq. \eqref{eq:boundai} to bound the first term, and  Eq. \eqref{eq:localresult} with $\vert B \vert = l^*$ to bound the second. With these, we conclude that there exists constants $c_1,c_2$ depending on all the constants involved (i.e. $\beta,h,J,k,c',v$) such that
\begin{equation} \label{eq:Zapprox}
     \left \vert   \tr[\rho_i A_i]- \tr[\rho_i^{(2l_*)} A^{l_*}_i]  \right \vert \le c_1 e^{-c_2 l^*}.
\end{equation}
The key feature of $\tr[\rho_i^{(2l_*)} A^{l_*}_i]$ is that it is an expectation value of an operator whose form we known explicitly, as per Eq. \eqref{eq:QBPoperator2}, evaluated in a thermal state of size $2l^*$. This can be computed exactly (or rather, with a subleading error) in a time $\text{exp} \left(\mathcal{O}(l^*)\right)$. Let us now choose a precision $\varepsilon/N$ in Eq. \eqref{eq:Zapprox}, so that $l^* = \mathcal{O}\left (\log{N/\varepsilon^{-1}} \right)$. This way, we have
\begin{align}
 &Z' \equiv  d^{N} \prod_{i=1}^{\vert E \vert} \tr[\rho_i^{(2l_*)} A^{l_*}_i] \\& = Z \prod_{i=1}^{\vert E \vert} \left (1+ \frac{\varepsilon}{N\tr[\rho_i A_i]}\right) = Z \left(1+ \mathcal{O}\left ( \varepsilon\right) \right).
\end{align}
The last equation comes from the fact that $N \propto \vert E \vert$ and that all eigenvalues of $A_i$ are $\mathcal{O}(1)$, as per the definition in Eq. \eqref{eq:QBPoperator}.
The algorithm thus consists of exactly calculating the numbers $\{\tr[\rho_i^{(2l_*)} A^{l_*}_i] \}$ exactly, and then multiplying them, so that
\begin{equation}
     \left \vert \log Z' - \log Z \right \vert \le \mathcal{O}\left ( \varepsilon\right).
\end{equation}
Since there are $ \vert E \vert \propto N$ terms in $Z'$, and each takes time $\text{poly}(N \times \varepsilon^{-1})$, the final runtime is $ \text{poly}(N,\varepsilon^{-1})$, as desired. See also \cite{Fawzi_2023} for a related result in the translation invariant setting.

%%%%%%%%%
%%%%%%%%%

\subsection{Hardness of approximating partition functions}\label{sec:Qpartition}

In the previous section we have seen how the partition function can be approximated in 1D in the sense of Eq. \eqref{eq:approxZ1D} as long as the temperature is $\beta = \mathcal{O}(1)$. Moreover, through the cluster expansion we briefly explained in Sec. \ref{sec:connectedclusters} how it can be approximated for any local model as long as $\beta < \beta^*$, where $\beta^*$ is some fixed constant \new{independent of system size.} 

On the other hand, in the limit of $\beta \rightarrow \infty$, the log-partition function equals the energy of the ground state. For classical models, approximating this to a certain precision is an NP-complete problem. For local quantum Hamiltonians, it is QMA hard. This means that there should be no efficient classical or quantum algorithm to approximate log-partition functions for low enough temperatures, both for classical and quantum models. In fact, it is known that the classical problem is only slightly harder than NP \cite{Stockmeyer1983Z} \footnote{More specifically, it is in the class $\text{BPP}^{\text{NP}}$, see \cite{bravyi2021complexity}.}, and that it is at least $\#P$ hard if complex interactions are allowed \cite{goldberg2017complexity}. For the quantum case, the exact complexity class to which this belongs or is complete for is not yet clear (see \cite{bravyi2021complexity} for more details and results). 

There is still the expectation that for certain classes of interesting models we can still compute the partition function efficiently, even with classical algorithms and at very low temperatures. One notable example are quantum Monte Carlo methods \cite{Bravyi2017,Crosson2021,crosson2020classical}, which are restricted to Hamiltonians without the so-called ``sign problem" (or \emph{stoquastic} \cite{Stoq2008}). Other results cover different specific kinds of models \cite{bravyi2021complexity,helmuth2022efficient}. Quantum algorithms for approximating general partition functions also exists \cite{Poulin2009,bravyi2021complexity,cade2017quantum,Chowdhury2021}, but often come with exponential run-times. 

\new{
Another relevant angle of this problem is the connection of efficient algorithms to the idea of completely analytical interactions from Sec. \ref{sec:completely}, as well as the physics of phase transitions. The intuition is that a physical phase transition in the system may come together with a \emph{computational} phase transition in which approximating $\log Z$ becomes fundamentally harder. Along these lines it has been shown that in quite a general setting \cite{Harrow_2020}, that the analiticity of the log-partition function implies the existence of an efficient algorithm. This can be understood in terms of the setting of Sec. \ref{sec:connectedclusters}: as long as the Taylor expansion converges well, we can compute the individual coefficients of Eq. \eqref{eq:logZ} efficiently.} %There are important open questions, however. %For instance we may expect that away from a phase transition, whenever the correlations decay exponentially, the partition function will be analytic and can be approximated efficiently.

%%%%%%%%%%
%%%%%%%%%%

\subsection{Tensor network methods}

Tensor network (TN) techniques are perhaps the most successful way of classically computing physical properties of quantum systems in 1D, and sometimes 2D \new{and often come with rigorous theoretical guarantees.} This includes most notably the regime of low energy physics \cite{verstraete2006matrix,Hastings2007,landau2013polynomialtime,Arad2017,huang2015polynomialtime} and, as we now review, that of finite temperature too. See e.g. \cite{Bridgeman_2017,eisert2013entanglement,Orus2019TN,Review2021} for introductory texts to this topics.% A detailed explanation of TN is beyond the scope of this tutorial, so we encourage the reader to first check the numerous introductory reviews on the subject e.g. . 

\begin{figure}[t]
\includegraphics[width=0.7\linewidth]{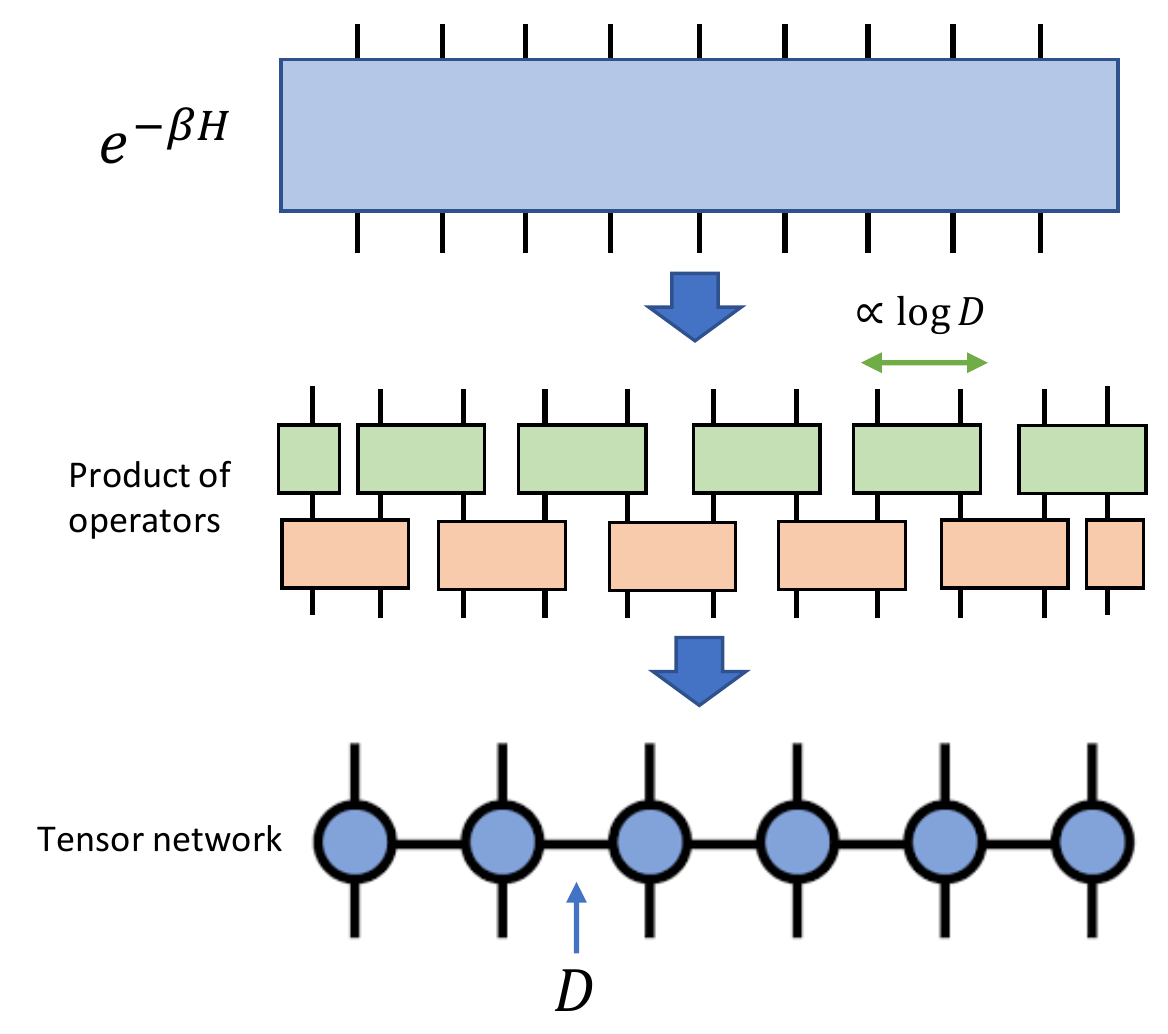}.
\caption{Schematically, the way to prove that a 1D thermal state is a tensor network is by decomposing it as a product of smaller operators. It then follows from standard methods that the bond dimension of the tensor network representation is related to the size of those operators.}\label{fig:chainTN}
\end{figure}

The \new{main} aim is to obtain a TN representation  of an operator $M_D$ such that $\vert \vert e^{-\beta H}- M_D \vert \vert_1 \le \varepsilon Z$, which then allows us to compute all thermal expectation values up to error $\varepsilon$ as per Eq. \eqref{eq:1norm}. The index $D$ labels the bond dimension which, roughly speaking, quantifies the complexity of representing $M_D$. A TN of bond dimension $D$ requires a memory $\propto N \times D^2$ to be stored. Intuitively, the approximation operator $M_D$ should be made out of a sum or low-depth product of operators with smaller support i.e. of size at most $\propto \log D$. This is graphically described for 1D in Fig. \ref{fig:chainTN}.

%Let us first briefly cover the best currently known result for 1D chains \cite{kuwahara2020}, in which the approximation is with the so-called Matrix Product Operators (MPO) \cite{Pirvu_2010}.

One possible way to do this is by adding piece by piece from right to left, aided by the results from Sec. \ref{sec:localityestimates}. To do this, first define segments of the chain of length $l$ such that the Hamiltonian up to segment $j-1$ is defined by $H_j$. This allows us to define the operator
\begin{equation}
    \Psi_j=e^{\beta H_j} e^{-\beta H_{j+1}},
\end{equation}
such that $e^{-\beta H}=  e^{-\beta h_1} \prod_{j=1} \Psi_j$. Each operator $\Psi_j$ can then be approximated by a localized operator $\Psi_j^l$ with support in a region of length $l$ as in Eq. \eqref{eq:araki2}, exponentially well in $l$. This can take the form
\begin{equation} \label{eq:psijl}
    \Psi^l_j=e^{\beta h_j} e^{-\beta h_{j+1}},
\end{equation}
There are \new{ $\mathcal{O} (N)$ } of those operators, and the error \new{$e^{-\Omega (l)}$} of each approximation can be shown to contribute additively. Thus, choosing $l \propto \log{N/\varepsilon}$ gives the desired $\epsilon$-good approximation to $e^{-\beta H}$. By construction, the product of operators resembles that of Fig. \ref{fig:chainTN}. In that case, the bond dimension can be straightforwardly assumed to be $D \le e^{\mathcal{O}(l)} = \text{poly}(N, \varepsilon^{-1})$, which is already computationally efficient \new{and likely close to optimal.}

A non-trivial improvement to this can be found in \cite{kuwahara2020}. Roughly speaking, one can define an operator $\tilde \Psi_j^l$ in which the exponential functions of Eq. \eqref{eq:psijl} are approximated by their Taylor series. In that case, we can put forward results about the bond dimension required to represent polynomials of Hamiltonians \cite{arad2013area}. This leads to an improvement of the bond dimension to $ D \le e^{ \tilde{\mathcal{O}}(\sqrt{l})}= \text{exp}\left( \tilde{\mathcal{O}}( \sqrt{\log(N/\varepsilon)}) \right)$. This is sub-linear in system size, much more computationally efficient. %This algorithm is in fact not very far from some that have been implemented in practice \cite{Chen2008}.

In higher dimensions, the best-known method is a variation of the cluster expansion proposed in \cite{hastings2006solving}. There, the expansion is treated in a slightly different way, to approximate the exponential $e^{-\beta H}$ rather than the log-partition function as in Sec. \ref{sec:connectedclusters}. Instead of counting the number of individual clusters of at most size $m$, one has to consider arbitrary products $\prod_{i \in W} h_i \equiv h(W)$ of terms $h_i$ from a multiset $W$ of size $\vert W \vert$, that can be divided into connected clusters. Let us label the multisets $W = \{ h_i\}$ in which the biggest cluster has size $M$ to be $C_M$. That this is consistent with the cluster expansion can be seen from taking the exponential of Eq. \eqref{eq:logZ} given the expression in terms of clusters of the powers in Eq. \eqref{eq:Kmoments}.

The following result was proven in detail in \cite{kliesch2014}. It reads
\begin{equation}\label{eq:clusterTN}
    \vert \vert e^{-\beta H} - \sum_{W \in C_M} \frac{(-\beta)^{\vert W \vert}}{\vert w \vert !} h(W) \vert \vert _1 \le Z \left(e^{N \frac{b(\beta)^M}{1-b(\beta)}}-1 \right), 
\end{equation}
where $b(\beta)<1$ for all $\beta < \beta^* = \mathcal{O}(1)$. In \cite{Molnar_2015} the sum over clusters on the RHS of Eq.~\eqref{eq:clusterTN} was shown to be a tensor network (in fact, a so-called PEPO) of bond dimension $e^{\mathcal{O}(M)}$. Thus, by setting the RHS to be $\varepsilon$, we achieve a TN approximation to $e^{-\beta H}$ with bond dimension $\text{poly}(N, \varepsilon^{-1})$. This only holds for inverse temperatures below $\beta^*$, but the result can be extended to arbitrary temperatures simply by taking powers of the operator. This means the bond dimension grows as $D \le \exp \left(\mathcal{O} \left( \beta \log{\frac{\beta N}{\varepsilon}} \right)\right)$ (see \cite{Molnar_2015} for the details). This scheme has recently been numerically implemented in practice \cite{vanhecke2021simulating}.

\new{These results might seem surprising, since they  show that there are in principle efficient TN representations for all dimensions and all temperatures $\beta =\mathcal{O}(1)$. This contradicts the intuition (justified by numerical works \cite{TN0,TN1,TN2,TN3,TN4,TN5}) that, at phase transitions, when long-range correlations are present, such efficient schemes should not exist. 
The caveat, however, is
 that in dimensions higher than one, a TN representation is not enough to be able to extract numerical data efficiently. This is because the contraction of TN can be a computationally demanding task by itself \cite{Schuch2007Comp,Haferkamp2020Contracting}. In fact, what we expect is that the ability to reliably contract a higher dimensional tensor network is related to facts such as local indistinguishability \cite{verstraete2004,Cirac_2013}, which allows us to obtain reliable results by contracting suitably smaller regions. }

Finally, let us note that by using the local indistinguishability from Sec. \ref{sec:localind} (or even without it in 1D \cite{HuangLocally}) it can be shown that a much smaller bond dimension is needed to simulate local properties \cite{alhambra2021locally} . 

%%%%%%%%%%%%%%
%%%%%%%%%%%%%%

\subsection{Quantum algorithms for preparing thermal states}\label{sec:Qalgorithms}

One of the most promising applications of quantum computers is the generation of exotic states of matter in complex many-body models. The expectation is that this should allows us to discover a potentially wide variety of physics, and also serve as a subroutine in certain quantum algorithms, such as those performing optimization tasks. 

Because of this, a question that has been very much explored lately is that of how to prepare thermal states of local Hamiltonians with a quantum computer. This could be either a fully fledged fault-tolerant one or one more suitable for the so-called NISQ (Noisy Intermediate Scale Quantum) devices. In the following we review some of the currently existing ones, and also explain the ideas that highlighting the complexity of the problem. We mostly focus on those that have some provable performance guarantees. There are many others we will not cover (such as e.g. \cite{Motta2020,Sirui2021,Cohn2020} and others), which often rely on some level of heuristic arguments. These also include approaches such as variational algorithms \cite{farhi2014quantum,chowdhury2020variational,Xinvariational2021,Cerezo20212} or those based on quantum versions of metropolis sampling \cite{Terhal2000,Temme2011,Yung2012}. These may nonetheless be more efficient in many physically relevant settings.

\subsubsection*{General considerations}

We have very strong evidence pointing that preparing thermal states is, in its most general setting, not an easy task. The results on QMA hardness of the local Hamiltonian problem \cite{QMA2005} show that there are vanishingly small temperatures (scaling quickly with system size) at which the preparation of $\rho_\beta$ is QMA complete. That is, not even a quantum computer can do it efficiently \cite{kitaev2002classical}. This is the case even for 1D systems \cite{Aharonov2009}.

\new{There are also reasons to believe that the problem is not easy even at slightly higher temperatures. For instance, it was recently shown \cite{anshu2022nlts} (following a famous conjecture \cite{HastingsPCP2013}) that there are local models for which preparing states below a certain energy density (including low temperature Gibbs states) requires a circuit of depth at least $\log N$. This property, however, does not apply to lattices \cite{AharonovPCP2013}.}

In that sense, we expect that large interesting classes of models and \new{temperature ranges} will have efficient algorithms. The locality of the model, and some of its consequences from the previous sections, should often simplify this task. 

%\alv{There are also compelling reasons to believe that this will be the case even at slightly higher temperatures: it has recently been shown that there exists local models for which all states below a certain energy density cannot be efficiently prepared with a quantum computer \cite{HastingsPCP2013}. The current best results along these lines are \cite{Eldar2019,AnshuNirkhe2022}, which show that there exist Hamiltonians for which the low energy thermal states (that is, with vanishingly small temperature) have a provably large circuit complexity lower bound. This takes the form of a lower bound in the number of elementary gates required to generate it. Stronger results also hold if one restricts the models and circuits to obey certain symmetries \cite{Bravyi2020}. It is then no surprise that the most general algorithms with a provable performance have a super polynomial (in fact, exponential) circuit complexity, even already at $\beta \sim \mathcal{O}(1)$.}

\subsubsection*{Algorithms based on purifications}

These algorithms work for general Hamiltonians, and \new{are often designed to} be run in a fully fault tolerant quantum computer, capable of applying any quantum circuit without large errors. They \new{aim to construct the following state
\begin{equation}
    \vert \rho_\beta \rangle = \frac{1}{\sqrt{Z}}\sum_{l} e^{-\beta E_l/2} \vert E_l \rangle_{A} \vert l \rangle_{\mathcal{A}},
\end{equation}
where the second subsystem $\mathcal{A}$ is made of auxiliary particles with an orthogonal basis $\{\vert l \rangle \}$ such that, upon tracing out, yield $\tr_{\mathcal{A}}[  \vert \rho_\beta \rangle \langle \rho_\beta \vert]= \rho_\beta$. The first stage of the algorithm involves preparing a state $\vert \psi \rangle$ with $ \vert \rho_\beta \rangle $ as a component such that 
\begin{equation}\label{eq:targetsup}
    \vert \psi \rangle = \frac{1}{\mathcal{N}} \vert \rho_\beta \rangle \vert 0 \rangle_R +....\quad ,
\end{equation}
where we have included a possible additional register $R$.}

\new{The first works proposing this scheme \cite{Poulin2009,chiang2010quantum,Bilgin2010} instead apply the phase estimation algorithm \cite{PhaseEstimation}. To a good approximation, this algorithm acts as follows \footnote{This output is stricly speaking only approximate, but the error of this approximation can be dealt with using additional registers (see \cite{Poulin2009})}
\begin{align}
U_{PE} ( \vert E_l \rangle_A  \vert 0 \rangle_{\mathcal{A}} ) = \vert E_l \rangle_A  \vert l \rangle_{\mathcal{A}},
\end{align}
that is, it ``measures" the energy of system $A$ into the register $\mathcal{A}$. Inputting a uniform superposition $\frac{1}{d^{N/2}} \sum_l \vert E_l \rangle$ yields
\begin{align}
U_{PE} ( \frac{1}{d^{N/2}} \sum_l \vert E_l \rangle\vert 0 \rangle_{\mathcal{A}} ) = \frac{1}{d^{N/2}} \sum_l \vert E_l \rangle\vert l \rangle_{\mathcal{A}}.
\end{align}
Now we add an additional qubit register on the state $\vert 0 \rangle_R$ and rotate it to $\vert \theta \rangle = \cos \theta \vert 0 \rangle + \sin \theta \vert 0  \rangle $ by an angle $\theta(E_l, \beta)=\arccos(e^{-\frac{\beta E_l}{2}})$ conditioned on the system, obtaining
\begin{align}\label{eq:finalalgo}
    \frac{1}{d^{N/2}} \sum_l \vert E_l \rangle\vert l \rangle_{\mathcal{A}} \vert \theta(E_l, \beta)\rangle_R = \sqrt{\frac{Z}{d^N}} \vert \rho_\beta \rangle \vert 0 \rangle_R +.... \quad
\end{align}
which is the target state with $\mathcal{N}= \sqrt{\frac{d^N}{Z}}$.
Other approaches use more recent quantum simulation ideas, such as the technique based on sums of unitaries  \cite{chowdhury2016quantum}, which leads to a better dependence on the approximation error $\epsilon$ in many cases of interest. }

\new{Finally, to obtain $\vert \rho_\beta \rangle$ with high precision, one must then apply amplitude amplification of the state $\vert \psi \rangle$, to output the component of Eq. \eqref{eq:finalalgo} corresponding to the register state $\vert 0 \rangle_R$. The gate complexity of this, however, grows linearly in $\mathcal{N}$, which sets the leading (almost) exponential gate cost of the algorithm. }

\new{There already exists improvements to this type of scheme.} In one dimension, one can instead implement this same algorithm connecting subsequent segments of the chain, which can reduce the gate complexity to a polynomial $\sim N^{\mathcal{O} (\beta)}$ \cite{Bilgin2010}. Also, recent progress shows that the phase estimation and amplitude amplification steps in these schemes can instead be replaced by random circuits with post-selection \cite{shtanko2021algorithms}, making them more amenable to current technologies. For commuting Hamiltonians, a purification in the form of a tensor network state (a PEPS) can be very efficiently prepared through an adiabatic algorithm \cite{Ge_2016}. See also the recent \cite{Holmes2022}, which produces a purification of a thermal state $\propto e^{-\beta H_1}$ starting from that of another Hamiltonian $H_0$, and is efficient when $\norm{H_0-H_1}$ is not too large. \new{A potentially efficient scheme along these lines is the one presented in \cite{Motta2020}.}

\subsubsection*{Efficient algorithms from physical features}

Perhaps the main caveat of most of the aforementioned algorithms is that they are constructed for very general Hamiltonians: they do not always make a very clear use of the physical features that we expect could simplify the problem, such as locality or any one of its consequences. \new{However, we expect that there exists efficient schemes to prepare Gibbs states of local model that belong to the class of ``completely analytical interactions", aided by facts such as decay of correlations.}

This is the case for the proposal in \cite{brandao2019finite}, whose efficiency depends on two such factors: the speed of decay of CMI from Sec. \ref{sec:CMI}, and the error in the local indistinguishability from Sec. \ref{sec:localind}. The algorithm uses iterations of the recovery map that appeared in Eq. \eqref{eq:recovery}, which are guaranteed to yield a low error if the CMI decays quickly enough. The main idea is that one can construct local decoupled parts of the thermal state independently, and then join them together to make up the whole $\rho_\beta$ via subsequent applications of the recovery map. The local indistinguishability guarantees that the local parts used in the recovery are also accurate parts of the whole thermal state the algorithm constructs. \new{The results on the exponential decay of correlations described in Sec. \ref{sec:corrdecay} and of exponential decay in CMI from Sec. \ref{sec:CMI} thus guarantee that there exists efficient algorithms for local models with a high enough temperature $\beta \le \beta^*_{\text{NC}}$, and also for 1D systems as long as the corresponding assumptions on the correlation decay are satisfied.} %A potential issue with this algorithm is that it requires the implementation of the recovery map, for which we do not always have explicit expressions \cite{Sutter2018}. It is encouraging, however, that quantum algorithms for one such recovery map, the Petz map, have appeared in the literature \cite{gilyen2020quantum}.

Another potential alternative route along these lines is to find out under which conditions the dissipative dynamics (that is, when the system is coupled weakly to some external bath) associated to a Gibbs state converge quickly. Then, tools to engineer dissipative dynamics can be in principle implemented in a quantum computer \cite{Kliesch2011,OpenSu2020}. The challenge is to find under which conditions these dynamics have a fast convergence or mixing rate. Rigorous results along these lines are so far mostly limited to  commuting Hamiltonians, as we explain in Sec. \ref{sec:commuting}.

%%%%%%%%%%%%%%%
%%%%%%%%%%%%%%%

\section{Commuting Hamiltonians}\label{sec:commuting}

There is a much simpler and yet physically relevant class of Hamiltonians that merits \new{a specific mention}: those in which all the $\{h_i\}$ commute with each other. This includes many interesting models for quantum many-body physics and quantum computation. It includes all stabilizer Hamiltonians, including the toric code and other widely studied examples, as well as many other models describing various topological phases of matter. 

Notice that these are not the same as classical Hamiltonians: even if we can diagonalize all the $h_i$ simultaneously, the energy eigenbasis will in general be highly entangled. In contrast, classical Hamiltonians have a product eigenbasis. At the same time, we have
\begin{equation}\label{eq:commuting1}
e^{-\beta (H - h_i)} = e^{-\beta H}e^{\beta h_i}=e^{\beta h_i/2}e^{-\beta H}e^{\beta h_i/2},
\end{equation}
so the tools in Sec. \ref{sec:localityestimates} and \ref{sec:QBP} are unnecessary. This means  that many of the results described above take much simpler forms and easier proofs, as we now briefly explain. 

Let us divide the lattice into two complementary regions $D,E$, with boundary $\partial_{DE}= \partial_D \cup \partial_E$, so that $H=H_D+H_E+H_I$, with $\supp{H_I} \in \partial_{DE}$. Notice that
\begin{equation}
    \tr_E[e^{-\beta H}]=e^{-\beta H_D} \tr_E[e^{-\beta (H_E+H_I)}].
\end{equation}
Clearly $\tr_E[e^{-\beta (H_B+H_I)}]$ has non-trivial support on the region  $\partial_D $ only. This means that the local indistinguishability from \ref{sec:localind} holds with no error by choosing $A= D $, $B=E \cap \partial_{DE}$, $C=E \setminus B$, so that $\text{dist}(A,C)$ is roughly the width of the boundary. A similar exact result applies to the Hamiltonian of mean force. We now briefly show the proof, which is elementary and can be found in \cite{anshu2021commuting}.  If we define $e^{-\beta \Phi} \equiv \tr_E[e^{-\beta (H_E+H_I)}]$, we see that 
\begin{equation}
    \frac{-1}{\beta}\log( \tr_E[e^{-\beta H}])= \alpha \mathbb{I}+ H_D + \Phi,
\end{equation}
where $\alpha$ is some constant, and $\Phi$ is localized in $D \cup \partial_{DE} $ and has bounded norm, as
\begin{equation}
    H_D+H_E - h \vert \partial_{DE} \vert \le H \le   H_D+H_E + h \vert \partial_{DE} \vert
\end{equation}
implies that 
\begin{equation}\label{eq:commutingdec}
    e^{-\beta h \vert \partial_{DE}\vert} e^{-\beta (H_D+H_E)} \le e^{-\beta H} \le  e^{\beta h \vert \partial_{DE} \vert} e^{-\beta (H_D+H_E)},
\end{equation}
which upon tracing $E$ out and multiplying by $e^{\beta H_D}$, implies that $\vert \vert \Phi \vert \vert \le 2 h \vert \partial_{DE}\vert$.

It should also be no surprise then that the Markov property of Sec. \ref{sec:CMI} also holds exactly. This means that if we define regions  $A,B,C$ such that $A,C$ are shielded by region $B$, we have that $I(A:C \vert B )=0$ \cite{PoulinLeifer2008,Poulin2011}. In fact, a converse statement holds (vanishing CMI implies the state is a thermal state of a local Hamiltonian) when the interaction graph $\Lambda$ is triangle-free \cite{brown2012quantum}. As mentioned in Sec. \ref{sec:CMI}, this is the quantum equivalent of the Hammersley-Clifford theorem \cite{hammersley1971markov}. 

All these exact results strongly suggest that algorithms such as those described in Sec. \ref{sec:algorithms} are much more efficient in this setting. For instance, it is immediate from a repeated application of Eq. \eqref{eq:commuting1} that the thermal states can be expressed exactly as tensor networks with constant bond dimension $D \le e^{\mathcal{O}(k)}$. There also exists quantum algorithms for commuting Hamiltonians that are significantly more efficient than the general ones in Sec. \ref{sec:Qalgorithms}  \cite{Ge_2016}. \new{In fact, the exact Markov property guarantees that Gibbs states of finite temperature can always be prepared efficiently (in linear time), simply by iterating applications of the Petz recovery map \cite{petz1986sufficient}.} %Also, further concentration inequalities akin to Eq. \eqref{eq:hoeffding} have been shown specifically for commuting Hamiltonians \cite{de2021quantum}.

%\alv{we might want to somewhat rewrite this}Due to their additional simplicity, there are also a number of ideas that have so far only been proven for these Hamiltonians. A noteworthy example are results on dissipative evolutions that map any given initial state to the thermal state. 

\new{Another important fact about commuting Hamiltonians is that, when weakly coupled to an external heat bath, the dissipative dynamics is known to remain local. This process is modeled by a Lindblad equation of the form}
\begin{equation}
    \frac{\text{d}\rho}{\text{d}t}=  \mathcal{L}(\rho)= -i[H,\rho]+ \sum_{\alpha} L_\alpha \rho L_\alpha^\dagger - \frac{1}{2}\left \{ L_\alpha L_\alpha^\dagger, \rho \right \},
\end{equation}
where $L_\alpha$ are local ``jump" operators and $\alpha$ indexes the energy gaps of $H$. The best known example are the Davies generators \cite{Davies1974}. See e.g. \cite{breuer2002theory,rivas2012open} for introductory references. 

\new{The interesting cases are those for which $\rho_\beta$ is the unique fixed point, such that $\mathcal{L}(\rho_\beta)=0$. The important question then is how long does this local dissipative evolution $e^{t \mathcal{L}}(\rho)$ take to approach the Gibbs state?} This can be tackled by analyzing the spectral gap and the log-Sobolev constant of $\mathcal{L}$. A bound on the spectral gap was proven assuming the decay of correlations in \cite{Kastoryano2016}, and for specific models in \cite{Alicki2009,Komar2016,Temme2017,lucia2021thermalization}. This shows that it takes $\text{poly}(N)$ time to thermalize. On the other hand, a bound on the so-called \emph{log-Sobolev contant} \cite{Kastoryano2013} instead constraints that to $\mathcal{O}(\log{N})$. This was recently proven for 1D chains \cite{bardet2021entropy,bardet2021rapid} and in \cite{bardet2021-heatbath,capel2020modified} for other models of dissipation. %These results show that the dissipative processes at hand can also be seen as an efficient quantum algorithms preparing thermal states, since in principle they can be simulated efficiently with a quantum computer \cite{Kliesch2011}.

%%%%%%%%%%%%%%%%
%%%%%%%%%%%%%%%%

\section{Conclusions and open questions}\label{sec:conclusion}

\new{There are many different models and systems for which we would like to know their properties at equilibrium. This is due to their pervasive presence in physics, but also due to their appearance in learning and sampling algorithms.}

It may appear at first that studying thermal states of general complex quantum models is a very challenging task. We hope to have illustrated the fact  that this is not always the case: for a large array of situations involving local Hamiltonians many non-trivial analytical statements can be made. These are both about universal physical features of the models at hand, but also about the computational complexity of the problems the physics poses. The connections found motivate a timely research program, largely inspired by quantum information theory: to understand the links between fundamental physical features and their computational complexity.

In the present context, much of the technical difficulty lies in working with with the matrix exponential of any such a Hamiltonian, in which typically the individual terms do not commute.  As seen in Sec. \ref{sec:tools}, however, we have a number of mathematical tools to deal with these in many physically relevant regimes.

\subsection{List of open questions}
We have covered a number of statements in different areas and summarized many of the existing results on the topic. However, plenty of relevant questions are still open. We now summarize some of them, which we believe to be of particular physical or technical interest:

\begin{itemize}

\item \new{In Sec. \ref{sec:connectedclusters} we have explained the technical concept of \emph{cluster expansion}, and their large number of applications in this context. Understanding its convergence further, in particular when considering the expansion of operators as done in \cite{Kuwahara_2020_Markov}, seems crucial for understanding relevant ideas such as the Hamiltonian of mean force (Sec. \ref{sec:meanforce}), and the decay of the conditional mutual information (Sec. \ref{sec:CMI}). It could also be interesting to extend them to long-range interacting systems \cite{Tran2021}.}

   \item The ideas of Sec. \ref{sec:subsystem}, and in particular the Hamiltonian of mean force, have in the past few years features in the study of thermodynamic quantities for strongly coupled systems \cite{MartiHenrik2018,Talkner2016,Strasberg2019,miller2018hamiltonian,talkner2020colloquium}. Many existing results on this topic focus on simpler models than those considered here, such as individual spins coupled to quadratic baths \cite{cresser2021,trushechkin2021open}. \new{In Sec. \ref{sec:strongcoupling} we have outlined how one can also answer thermodynamic questions about strongly coupled spin systems. It would be interesting to further explore whether the results from Sec. \ref{sec:subsystem} have further non-trivial consequences, such as those found in \cite{AlhambraFundamental2019}.}
   
 %  \item The current theoretical results for tensor network descriptions of thermal states give reasonably good bounds on the bond dimension \cite{kuwahara2020,Molnar_2015,kliesch2014}. It would be interesting, however, to know whether the tensor networks in dimensions higher than one \cite{kliesch2014,Molnar_2015} can be contracted efficiently to calculate expectation values, perhaps in cases for which the Gibbs state has exponential clustering of correlations (see \cite{Cirac_2013,Schwartz2017} for similar ideas for ground states). A very interesting complementary question is to fully understand whether the set of density operators represented by tensor networks in 1D is contained within the set of thermal states of local Hamiltonians \cite{chen2020matrix}.
   
   \item With the advent of quantum computing, there are multiple ongoing efforts aiming to find more efficient quantum algorithms for thermal sampling and partition functions. As we have seen in Sec. \ref{sec:Qalgorithms}, many of the existing ones are designed for very general situations, and as such have performance bounds that will often be too conservative. Some existing schemes make use of relevant physical features to simplify them \cite{Bilgin2010,Ge_2016,brandao2019finite,chen2021fast}, but it seems that there is still plenty of room for exploring the kinds of regimes in which explicit and efficient algorithms can be proven. Since preparing thermal states is presumably an easier task than a general quantum computation (at least in certain regimes), it may be possible to tailor them to the limited capabilities of near-term noisy devices \cite{shtanko2021algorithms,zhang2023dissipative}. 
   
   \item \new{An important question when dealing with large quantum systems is to construct efficient ways to verify and characterize them. In the present context, the question is that of the complexity of the problem of thermal state tomography \cite{Anshu_2021,anshu2021commuting,haah2021optimal,rouze2021learning}.} The basic question is: can we learn the Hamiltonian from a small number of simple (local) measurements of few copies of $e^{-\beta H}/Z$? Optimal sample and computational complexity bounds exists in the high temperature regime, in which the cluster expansion applies \cite{haah2021optimal}, but beyond that our theoretical understanding is not complete (for instance, in 1D). This problem has a number of applications, including the verification of quantum computation in which thermal sampling is involved \cite{BrandaoSvore2017,brando_et_al:LIPIcs:2019:10603,vanApeldoorn2020quantumsdpsolvers,GSLBrandao2022fasterquantum}, or the characterization of many-body entanglement \cite{Kokail2021,Kokail2021-2}.
\end{itemize}

%%%%%%%%%%%%
%%%%%%%%%%%%

\section*{Acknowledgements}

AMA would like to thank the organizers of the Quantum Thermodynamics Summer School 2021, Nuriya Nurgalieva and L\'idia del Rio, since it triggered the idea of this tutorial. AMA also emphatically thanks the many colleagues and collaborators from which I have learnt about these topics through the years. Particular thanks go to Anurag Anshu and Tomotaka Kuwahara for useful comments and discussions, and to the authors of \cite{CapelQBP} for help with Sec. \ref{sec:QBP}. AMA acknowledges support from the Alexander von Humboldt foundation, the Centro de Excelencia Severo Ochoa
Program SEV-2016-0597 and the Ram\'on
y Cajal program RyC2021-031610-I , financed by
MCIN/AEI/10.13039/501100011033 and the European
Union NextGenerationEU/PRTR.

%%%%%%%%%%%%%%%%%%%%
%%%%%%%%%%%%%%%%%%%%

\appendix

\section{Derivations of Gibbs states}

\subsection{Gibbs weights from the ultra-weak coupling assumption}\label{app:weak}

\new{Here we sketch the standard derivation of how the Gibbs factor appears when a system is weakly coupled to a bath. Let us take a system-bath Hamiltonian
$H=H_S+H_B+ H_I$,
in which the interaction $H_I$ is arbitrarily weak. In that limit, we can approximate
\begin{equation}
    H=\sum_{E_S^{(j)}+E_B^{(i)}=E} E \, \vert E_s \rangle \langle E_s \vert \otimes \vert E_B^{(i)} \rangle \langle E_B^{(i)} \vert ,
\end{equation}
so that the eigenstates are product between system and bath.}

\new{A common and often relevant assumption is that the dynamics is ergodic, in the sense that we can describe the system-bath by the microcanonical ensemble, where all configurations of the same energy $E$ have equal probability. This is
\begin{equation}
    \frac{\Pi_E}{d_E} =  \frac{1}{d_E}\sum_{E_S^{(j)}+E_B^{(i)}=E} \vert E_S^{(j)} \rangle \langle E_S^{(j)} \vert \otimes \vert E_B^{(i)} \rangle \langle E_B^{(i)} \vert.
\end{equation}
The bath is typically understood as an infinitely large system, with an unbounded heat capacity $C = \frac{d \langle H_B \rangle_\beta}{d T}= -\beta^2 \frac{d \langle H \rangle_\beta}{d \beta}$. The bath also obeys the very weak constraint that its entropy is extensive with system size. Both these facts translate into the density of states of the bath $B$ having the following exponential form (see e.g. \cite{Richens_2018})
\begin{equation}
    \#(E_B) \propto e^{\beta E_B}.
\end{equation}
That is, the number of bath eigenstates $\vert E_B^{(i)} \rangle$ with energy $E_B$ is exponential in that energy.}

\new{We can use this to obtain the expression for the reduced density matrix on the system
\begin{align}
\tr_B[ \frac{\Pi_E}{d_E}] &\propto \sum_j \# (E-E_S^{(j)}) \vert E_S^{(j)} \rangle \langle E_S^{(j)} \vert \\& \propto \sum_j e^{-\beta E_S^{(j)}} \vert E_S^{(j)} \rangle \langle E_S^{(j)},
\end{align}
which are exactly the Gibbs weights.}

\subsection{Jaynes' maximum entropy principle}
\label{app:jaynes}

\new{
A well-known property that uniquely characterizes thermal states is the so-called maximum entropy principle. This specifies that of all the states with a given energy (or the expectation value of some other quantity) they are the state of largest possible entropy. To see this, let us choose $\rho \neq \rho_\beta$ such that $\tr[\rho H]=\tr[\rho_\beta H]$. Then,
\begin{align}
    S(\rho_\beta)-S(\rho) &= \tr[\rho\log \rho] +\beta \tr[\rho_\beta H]+ \log Z
    \\ &=  \tr[\rho\log \rho] +\beta \tr[\rho H]+ \log Z
    \\&=  \tr[\rho\log \rho] - \tr[\rho\log \rho_\beta] 
    \\& = D(\rho \vert \vert \rho_\beta) > 0.
\end{align}
Notice that these steps are unchanged if instead of considering just the Hamiltonian $H$ we take into account a higher number of charges $Q_i$ with their chemical potentials $\mu_i$, and the state $\exp(-\sum_j \mu_j Q_j) /\tr[\exp(-\sum_j \mu_j Q_j)]$.}

\new{
This simple principle is often interpreted as follows: if there is some state of which we only have partial information (in this case, its average energy), it is very often a good guess to assume it is the thermal state of that energy. Since it is the state with maximum entropy (which we can associate with ``maximum ignorance"), its choice makes the fewest assumptions about the structure of the actual state at hand. This idea is often applied in fields like statistical inference and optimization problems,\new{ as well as certain quantum algorithms} \cite{BrandaoSvore2017,brando_et_al:LIPIcs:2019:10603,vanApeldoorn2020quantumsdpsolvers,GSLBrandao2022fasterquantum}. It can also be seen as a variational definition that uniquely singles out thermal states. This allows for the application of this principle in different types of algorithms for finding or characterizing them \cite{Anshu_2021,DiGiorgio2021}.}

\section{Miscellaneous proofs}

\subsection*{Locality of operator $E_A$}\label{app:localproof}

In Sec. \ref{sec:localityestimates} we defined the operator
\begin{equation}
    E_A = e^{-\beta (H+A)}e^{\beta H}= \mathcal{T} e^{-\int_0^\beta \text{d}s e^{-s H} A e^{s H}},
\end{equation}
which is the solution of the differential equation
\begin{equation}
    \frac{\text{d}E_A}{\text{d}\beta} = -E_A A( i\beta),
\end{equation}
with $A( i\beta)=e^{-\beta H} A e^{\beta H}$. We can also define the localized generator
\begin{equation}
    A^l(i\beta)=\sum_{m=0}^{l} \beta^m C_m(A),
\end{equation}
and also the corresponding operator $E_A(l)$ as the solution of
\begin{equation}
  \frac{\text{d}E_A(l)}{\text{d}\beta} =- E_A(l) A^l(i\beta).
\end{equation}
Now from the Trotter-Suzuki decomposition
\begin{align}
    E_A= \lim_{L \rightarrow \infty} \prod_{j=0}^{L-1} e^{-A(i\frac{\beta j}{L})\frac{\beta}{L}} \\
     E_A(l)= \lim_{L \rightarrow \infty} \prod_{j=0}^{L-1} e^{-A^l(i\frac{\beta j}{L})\frac{\beta}{L}},
\end{align}
\new{we have that
\begin{align}
    &E_A -  E_A(l) = \lim_{L \rightarrow \infty} \sum_{j=0}^{L-1} \left(\prod_{j'=0}^{j-1} e^{-A^l(i\frac{\beta j'}{L})\frac{\beta}{L}}\right) \\ & \left( A^l(i\frac{\beta j}{L})\frac{\beta}{L}-A(i\frac{\beta j}{L})\frac{\beta}{L}  \right) \left(\prod_{j'=j+1}^{L-1} e^{-A(i\frac{\beta j'}{L})\frac{\beta}{L}}\right) \nonumber % \\ &=\int_0^\beta A^l(i s) - A(i s)\text{d}s.
\end{align}
Considering that Eq. \eqref{eq:normAbeta} also applies to the generator $A^l(i \beta)$,  using  \eqref{eq:normAbeta2} and \eqref{eq:normEa} and the triangle inequality repeatedly yields
\begin{align}
    \vert \vert E_A -  E_A(l) \vert \vert & \le \norm{E_A} \int_0^\beta \vert \vert A^l(i s) - A(i s) \vert \vert \text{d}s \\& \le \beta k \vert \vert A \vert \vert  \frac{(2 \beta Jk)^{l+1}}{(1- 2 \beta Jk)^{\frac{\norm{A}}{2\beta J}+1}}.
\end{align}}

\subsection*{Proof of Quantum Belief Propagation Eq. \eqref{eq:derivativeQBP}} \label{app:QBP}

\new{
The aim of this section is to give an expression for the derivative of the matrix exponential $\frac{\text{d} e^{-\beta H(s)}}{\text{d} s}$, where we assume $H(s)=H+s A$. These steps are elementary and have been omited in some previous relevant references \cite{Kim2012,Kato_2019,Anshu_2021}, but here we reproduce them in full as they appear in \cite{CapelQBP}
First, using DuHamel's identity, we can write
\begin{equation}
    \frac{\text{d} e^{-\beta H(s)}}{\text{d} s} = -\beta \int_0^1 e^{-\beta \tau H(s)}A e^{-\beta (1-\tau) H(s)} \text{d}\tau.
\end{equation}
We now expand the operator $A$ in the eigenbasis of $H(s)= \sum_i E_i(s) \vert i(s) \rangle \langle i(s) \vert$ as $A=\sum_{i,j} A_{i,j} \vert i(s) \rangle \langle j(s) \vert$, and write
\begin{align}\label{eq:QBPDer}
     & \frac{\text{d} e^{-\beta H(s)}}{\text{d} s} \\ \nonumber &= -\beta \sum_{i,j} A_{i,j}  \int_0^1 e^{-\beta \tau H(s)}  \vert i(s) \rangle \langle j(s) \vert e^{-\beta (1-\tau) H(s)} \text{d}\tau
      \\ \nonumber &=-\beta \sum_{i,j} A_{i,j}  \int_0^1 e^{\beta \tau \Delta E_{i,j}}  \vert i(s) \rangle \langle j(s) \vert e^{-\beta H(s)} \text{d}\tau
      \\ \nonumber &=-\beta \sum_{i,j} A_{i,j} (1+e^{\beta \Delta E_{i,j}} )^{-1}   \\ \nonumber &  \times \int_0^1 e^{\beta \tau \Delta E_{i,j}} \text{d}\tau \left \{ e^{-\beta H(s)},\vert i(s) \rangle \langle j(s) \vert \right\}
      \\ \nonumber & = -\frac{\beta}{2} \left \{ e^{-\beta H(s)}, \Phi_\beta^{H(s)(A)} \right\},
\end{align}
where $\Delta E_{i,j}=E_j(s)-E_i(s)$ and we define the operator
\begin{align}
    \Phi_\beta^{H(s)(A)}&= \sum_{i,j} \hat{f}_\beta(\Delta E_{i,j}) A_{i,j} \vert i(s) \rangle \langle j(s) \vert \\ & = \int_{-\infty}^{\infty} \text{d}t f_\beta(t) e^{-itH(s)} A e^{itH(s)},
\end{align}
where it can be seen from Eq. \eqref{eq:QBPDer} that the function $\hat{f}_\beta(\omega)$ is
\begin{equation}
    \hat{f}_\beta(\omega)= \frac{2}{1+e^{\beta \omega}}\int_0^1 e^{\beta \tau \omega}\text{d}\tau =\frac{2}{\beta \omega}\frac{e^{\beta \omega}-1}{e^{\beta \omega}+1},
\end{equation}
with its corresponding Fourier transform (as derived in Appendix B of \cite{anshu2021area})
\begin{equation}
    f_\beta(t)= \frac{2}{\beta \pi} \log \left( \frac{e^{\pi \vert t\vert/\beta}+1}{e^{\pi \vert t\vert/\beta}-1} \right).
\end{equation}
It is important to note that
\begin{equation} \label{eq:normft}
\int_{-\infty}^{\infty} f_\beta(t) \text{d}t =1
\end{equation}
and that, since by the fact that $\log x \le x-1$,
\begin{equation}
f_\beta(t) \le  \frac{4}{\beta \pi} \frac{1}{e^{\pi \vert t\vert/\beta}-1}.
\end{equation}
So both this function and its integral are exponentially small, in the sense that, for $a>\beta/\pi$,
\begin{align} \label{eq:ftdecays}
\int_{a}^\infty \text{d}t  f_\beta(t) &\le \frac{4}{\beta \pi} \int_{a}^\infty \text{d}t \frac{1}{e^{\pi \vert t\vert/\beta}-1} \\ \nonumber & = \frac{4}{\beta \pi (e^{\pi a/\beta}-1)} \int_{a}^\infty \text{d}t \frac{e^{\pi a/\beta}-1}{e^{\pi \vert t\vert/\beta}-1}
\\ \nonumber & \le \frac{4}{\beta \pi (e^{\pi a/\beta}-1)} \int_{a}^\infty \text{d}t e^{\pi (a-t)/\beta} \\ \nonumber & \le \frac{4}{\pi^2 (e^{\pi a/\beta}-1)}.
\end{align}
}
\subsection*{Proof of Eq. \eqref{eq:mineq2}}\label{app:mineq2}

This can also be found in \cite{Lenci2005}.
Let $F(t)$ be a differentiable and bounded operator. DuHamel's identity for a general operator function $F(t)$ states that
\begin{equation}
\frac{\text{d}}{\text{d}t}e^{F(t)}= \int_0^1 \text{d} u e^{u F(t)}\frac{\text{d}F(t)}{\text{d}t} e^{(1-u) F(t)}  . 
\end{equation}
Then we have that 
\begin{align}
    &\frac{\text{d}}{\text{d}t} \log \tr \left( C e^{H_1+t H_2} \right) \\&= \frac{\tr \left(\int_0^1 \text{d}u C e^{u (H_1+t H_2)}H_2 e^{(1-u) (H_1+t H_2)} \right)}{\tr \left( C e^{H_1+t H_2} \right)}\\ &= \frac{\tr \left(C'\int_0^1 \text{d}u  e^{(u-1/2) (H_1+t H_2)}H_2 e^{(1/2-u) (H_1+t H_2)} \right)}{\tr \left(C' \right)}\\  & \le  \left \vert \left \vert \int_0^1 \text{d}u  e^{(u-1/2) (H_1+t H_2)}H_2 e^{(1/2-u) (H_1+t H_2)} \right \vert \right \vert,\label{eq:proofC1}
\end{align}
where $C'=e^{\frac{H_1+t H_2}{2}}C e^{\frac{H_1+t H_2}{2}}$.
This follows from H\"older's inequality Eq. \eqref{eq:Holder} and the positivity of $C,C'$. Finally,
\begin{align}
  &   \left \vert \log \tr[C e^{H_1+H_2}]- \log \tr[C e^{H_1}] \right \vert \\ &= \left \vert \int_0^1 \frac{\text{d}}{\text{d}t}\log \tr \left[ C e^{H_1 + t H_2} \right ] \text{d}t  \right \vert
     \\ &\le \int_{0}^1 \text{d}t \int_{-1/2}^{1/2} \text{d}s \norm{e^{s(H_1+tH_2)}H_2 e^{-s(H_1+tH_2)}},
\end{align}
where the last step follows from the triangle inequality, Eq. \eqref{eq:proofC1} and the change of variable $u-1/2 =s$.

%%%%%%
%%%%%%

\subsection*{Proof of Eq. \eqref{eq:moments}} \label{app:usefullema}

This can also be found in \cite{Kuwahara_2020_ETH}.
Let $p(x)$ be an arbitrary probability distribution with $\int_{-\infty}^\infty x p(x) \text{d}x=a$, and the condition that $p(x)$ be Lebesgue integrable. We aim to bound
\begin{align}
&\int_{-\infty}^\infty \vert x-a\vert^m p(x) \text{d}x= \int_{-\infty}^\infty \vert x\vert^m p(x+a) \text{d}x \\ &=\int_{0}^\infty \vert x\vert^m (p(x+a)+p(-x+a)) \text{d}x\\&= -\int_0^\infty x^m \frac{\text{d}}{\text{d}x} \left [ \int_{\vert x'-a \vert \ge x} p(x') \text{d}x \right]\text{d}x,
\end{align}
where in the last step we used the fundamental theorem of calculus. This can now be integrated by parts as 
\begin{align}
  &  -\int_0^\infty x^m \frac{\text{d}}{\text{d}x} \left [ \int_{\vert x'-a \vert \ge x} p(x') \text{d}x' \right]\text{d}x \\&=- \left(x^m\int_{\vert x'-a \vert \ge x} p(x') \text{d}x' \right) \Bigg \vert_{0}^\infty \\ &+\int_{0}^\infty m x^{k-1} \int_{\vert x'-a \vert \ge x} p(x') \text{d}x' \text{d}x
    \nonumber \\ & \le \int_{0}^\infty m x^{k-1} 2 e^{-\frac{x^2}{4 c \bar A}} \text{d}x =(4 c \bar A)^{m/2} \left(\frac{m}{2}\right)!,
\end{align}
where in the second line the first term vanishes by definition, and in the third line we used the concentration bound Eq. \eqref{eq:hoeffding}.

\bibliography{references}

\end{document}